\RequirePackage{ifpdf}
\ifpdf 
\documentclass[pdftex]{sigma}
\else
\documentclass{sigma}
\fi

\newcommand\commentout[1]{}
\newcommand\boldB{{\boldsymbol{B}}}
\newcommand\boldL{{\boldsymbol{L}}}
\newcommand\boldW{{\boldsymbol{W}}}
\newcommand\calA{\mathcal{A}}
\newcommand\calF{\mathcal{F}}
\newcommand\calV{\mathcal{V}}
\newcommand\Complex{\mathbb{C}}
\newcommand\der{\partial}
\newcommand\mL{\mathcal{L}}
\newcommand\pa{\partial}
\newcommand\B{\mathcal{B}}
\newcommand\M{\mathcal{M}}
\newcommand\F{\mathcal{F}}
\newcommand\mP{\mathcal{P}}
\newcommand\Flag{\mathit{Flag}}
\newcommand\Integer{{\mathbb Z}}
\newcommand\mkp{{\mathrm{mKP}}}
\DeclareMathOperator{\ord}{ord}
\DeclareMathOperator{\Res}{Res}
\newcommand\symbolh{\sigma^\hbar}
\newcommand\SGM{SGM}
\newcommand\tensor{\otimes}

\numberwithin{equation}{section}

\newcommand\propref[1]{Proposition~{\rm \ref{#1}}}
\newcommand\secref[1]{Section~\ref{#1}}
\newcommand\appref[1]{Appendix~\ref{#1}}
\newcommand\corref[1]{Corollary~\ref{#1}}
\newcommand\lemref[1]{Lemma~\ref{#1}}
\newcommand\remref[1]{Remark~\ref{#1}}

\DeclareMathOperator{\ad}{ad} \DeclareMathOperator{\dmKP}{dmKP}
\DeclareMathOperator{\dcmKP}{dcmKP}

\begin{document}
\allowdisplaybreaks

\renewcommand{\PaperNumber}{072}
\renewcommand{\thefootnote}{$\star$}

\FirstPageHeading

\ShortArticleName{Coupled Modif\/ied KP Hierarchy and Its Dispersionless Limit}

\ArticleName{Coupled Modif\/ied KP Hierarchy\\ and Its Dispersionless Limit\footnote{This paper is a contribution 
to the Vadim Kuznetsov Memorial Issue ``Integrable Systems and Related Topics''.
The full collection is available at 
\href{http://www.emis.de/journals/SIGMA/kuznetsov.html}{http://www.emis.de/journals/SIGMA/kuznetsov.html}}}

\Author{Takashi TAKEBE~$^\dag$ and Lee-Peng TEO~$^\ddag$}
\AuthorNameForHeading{T. Takebe and L.-P. Teo}

\Address{$^\dag$~Department of Mathematics, Ochanomizu University,\\
$\phantom{^\dag}$~Otsuka 2-1-1, Bunkyo-ku, Tokyo, 112-8610, Japan} 

\EmailD{\href{mailto:takebe@math.ocha.ac.jp}{takebe@math.ocha.ac.jp}} 

\Address{$^\ddag$~Faculty of Information Technology, Multimedia University,\\
$\phantom{^\ddag}$~Jalan Multimedia, Cyberjaya, 63100, Selangor Darul Ehsan, Malaysia}

\EmailD{\href{mailto:lpteo@mmu.edu.my}{lpteo@mmu.edu.my}}

\ArticleDates{Received August 18, 2006, in f\/inal form October 03,
2006; Published online October 31, 2006}

\Abstract{We def\/ine the coupled modif\/ied KP hierarchy and its
dispersionless limit. This integrable hierarchy is a generalization of
the ``half'' of the Toda lattice hierarchy as well as an extension of
the mKP hierarchy. The solutions are parametrized by a f\/ibered f\/lag
manifold. The dispersionless counterpart interpolates several versions
of dispersionless mKP hierarchy.}

\Keywords{cmKP hierarchy; f\/ibered f\/lag manifold; dcmKP hierarchy}

\Classification{37K10} 

\begin{flushright}
\it Dedicated to the memory of Vadim Kuznetsov.
\end{flushright}

\section{Introduction}
\label{sec:intro}

Since 1980's many integrable systems with inf\/initely many degrees of
freedom have been studied by means of inf\/inite dimensional homogeneous
spaces. Well-known examples are: the KP hierar\-chy and the Sato--Grassmann
manifold (\cite{sat:81,sat-sat:82,sat-nou:84,djkm} etc.), the Toda lattice hierarchy and ``$GL(\infty)$''
(\cite{uen-tak:84,taka:84,tak:91a,tak:91b} etc.), the modif\/ied KP
(mKP) hierarchy and the f\/lag manifold (\cite{kas-miw:81,kac-pet:86,dic:99,kup:00,tak:02} etc.). In
this paper we add one more example to this series: the {\em coupled
modified KP} (cmKP) hierarchy and the f\/ibered f\/lag manifold.

The modif\/ied KP hierarchy  is def\/ined in \cite{dic:99} and
\cite{tak:02}
as a system consisting of two sets of equations: the Lax equations for
continuous variables $t=(t_1,t_2,\dots)$ and a set of dif\/ference
equations for the discrete variable $s$. The cmKP hierarchy has the same
description but the normalization of the operators
    is dif\/ferent. By
this dif\/ference the moduli space of solutions of the mKP hierarchy ($=$
the f\/lag manifold) is enlarged.

Actually a special case of the cmKP hierarchy has been known since
\cite{uen-tak:84}, in which the Toda lattice hierarchy was introduced. A
half of the Toda lattice hierarchy without dependence on half of time
variables is a cmKP hierarchy (See \appref{app:diff-ce}). Therefore the
cmKP hierarchy can be considered as the mKP hierarchy coupled to the
Toda f\/ield.

In this special case the solution space is parametrized by the basic
af\/f\/ine space $GL(\infty)/N$ where $N$ is the subgroup of inf\/inite upper
triangular matrices with unity on the diagonal. In other words it is a
product of the full f\/lag manifold and $(\Complex^\times)^\Integer$. The
solution space of our cmKP hierarchy is (partial f\/lag manifold) $\times
\prod\limits_{s\in S'}(\Complex^{m_s}\setminus \{0\})$ in general. (See
\corref{cor:sol<->(phi,flag)}.)

The dispersionless (quasi-classical) limit of the cmKP hierarchy is
taken in the same way as the dispersionless KP and Toda hierarchies. (See
\cite{TT1} and references therein.) We call the resulting system the
{\em dispersionless cmKP hierarchy} (the {\em dcmKP hierarchy} in
short).  In fact the dcmKP hierarchy was f\/irst introduced by one of the
authors \cite{teo:03} (in a slightly dif\/ferent form) as a system which
interpolates two versions of the dispersionless mKP hierarchies, one by
\cite{Kuper} and~\cite{CT} and the other by~\cite{tak:02}. Hence the
name ``dispersionless coupled mKP hierarchy'' has another
interpretation: It connects variants of the dispersionless mKP
hierarchy.

This paper is organized as follows:
The part on the cmKP hierarchy (\secref{sec:cmkp}) follows standard
recipe. We start from the Lax representation similar to that of the mKP
hierarchy~\cite{tak:02} and introduce the dressing operator and the
wave function as solutions of linear problems. We show existence of the
$\tau$ functions and construct them explicitly, using the DJKM free
fermions.

The dispersionless counterpart is discussed in \secref{sec:dcmkp},
following the strategy of \cite{TT1}. The basic objects are a formal
power series $\mL$ and a polynomial $\mP$. The hierarchy is def\/ined by
the Lax equations. Then we introduce the dressing function, the
Orlov--Schulman function, the $S$ function and the $\tau$ function. We
also discuss the relation with the dispersionless mKP hierarchy and the
characterization of the $\tau$ function.

In \secref{subsec:cmkp->dcmkp} the cmKP hierarchy and the dcmKP
hierarchy, so far discussed independently in principle, are related via
the WKB analysis.

Equivalent formulations of the cmKP hierarchy are discussed in the
appendices.

\renewcommand{\thefootnote}{\arabic{footnote}}
\setcounter{footnote}{0}

\section[Coupled modified KP hierarchy]{Coupled modif\/ied KP hierarchy}
\label{sec:cmkp}

\subsection[Definition of the cmKP hierarchy]{Def\/inition of the cmKP hierarchy}
\label{subsec:def-cmkp}

In this section we def\/ine the {\em cmKP hierarchy} with discrete
parameters $\{n_s\}_{s\in S} \subset \Integer$, where $S$ is a set of
consecutive integers (e.g., $S=\Integer$, $S=\{0,1,\dots,n\}$ etc.) as
in \cite{tak:02}. The dispersionless limit can be taken only when
$S=\Integer$, $n_s = Ns$. Set $S' = S \setminus 
\{\text{maximum element
of $S$}\}$ if there exists a maximum element of $S$ and $S'=S$
otherwise.

The independent variables of the cmKP hierarchy are the discrete
variable $s \in S$ and the set of continuous variables
$t=(t_1,t_2,\dots)$. The dependent variables are encapsulated in the
following operators with respect to $x$:
\begin{gather}
    L(s;x,t) = \der + u_1(s,x,t) + u_2(s,x,t) \der^{-1} + \cdots
\label{def:L(s)}
\\
\phantom{L(s;x,t)}{}= \sum_{n=0}^\infty u_n(s,x,t) \der^{1-n},
\nonumber
\\
    P(s;x,t) = p_0(s,x,t) \der^{m_s} + \cdots + p_{m_s - 1}(s,x,t) \der
\label{def:P(s)}
\\
\phantom{P(s;x,t)}{}= \sum_{n=0}^{m_s-1} p_n(s,x,t) \der^{m_s - n}.
\nonumber
\end{gather}
where $\der=\der_x$, $u_0=1$, $p_0\neq 0$, $m_s:=n_{s+1}-n_s$.
$P(s;x,t)$ is def\/ined only for $s\in S'$. We often write $L(s)$, $P(s)$
instead of $L(s;x,t)$, $P(s;x,t)$. The notation $(L(s),P(s))_{s\in S}$
stands for a pair of sequences $((L(s))_{s\in S}, (P(s))_{s\in S'})$.

The cmKP hierarchy is the following system of dif\/ferential and
dif\/ference equations:
\begin{gather}
    \frac{\der L(s)}{\der t_n} = [B_n(s), L(s)],
\label{cmkp:dL/dt}
\\
    L(s+1) P(s) = P(s) L(s),
\label{cmkp:LP=PL}
\\
    \left(\frac{\der}{\der t_n} - B_n(s+1) \right) P(s)
    =
    P(s) \left(\frac{\der}{\der t_n} - B_n(s) \right),
\label{cmkp:BP=PB}
\end{gather}
where $B_n(s) = B_n(s;x,t) = \bigl(L(s;x,t)^n\bigr)_{>0}$. The
projections like $(\cdot)_{>0}$ are def\/ined as follows: for
$A(x,\der) = \sum\limits_{n\in\Integer} a_n(x)\der^n$,
\begin{gather*}
    A_{>0} := \sum_{n>0} a_n(x) \der^n,\qquad 
    A_{\geq 0} := \sum_{n \geq 0} a_n(x) \der^n,
\\
    A_{<0} := \sum_{n<0} a_n(x) \der^n,\qquad 
    A_{\leq 0} := \sum_{n\leq 0} a_n(x) \der^n.
\end{gather*}
The last equation \eqref{cmkp:BP=PB} can be written in the form
\begin{gather}
    \frac{\der P(s)}{\der t_n} = B_{n}(s+1)\, P(s) - P(s)\, B_n(s),
\label{cmkp:dP/dt}
\end{gather}
as well. Since $B_1(s)=\der$, equations \eqref{cmkp:dL/dt} and
\eqref{cmkp:dP/dt} for $n=1$ imply that $x$ and $t_1$  always appear
in the combination $x+t_1$.

Note that the cmKP hierarchy is almost the same as the mKP hierarchy in
\cite{dic:99} or \cite{tak:02} but the forms of $L(s)$, $P(s)$ and
$B_n(s)$ are dif\/ferent.

\begin{remark}
\label{rem:P(s)}
 We can start from $P(s)$ with the $0$-th order terms, but if
 $(L(s),P(s))_{s\in\Integer}$ satisf\/ies the cmKP hierarchy, we can gauge
 away such terms. See \appref{app:P(s)} for details.
\end{remark}

By the well-known argument (cf.\ \cite[\S~1]{djkm}, \cite[Theorem 1.1]{uen-tak:84}) 
we can prove that the Lax equations \eqref{cmkp:dL/dt}
is equivalent to the Zakharov--Shabat (or zero-curvature) equations,
\begin{gather}
    \left[
    \frac{\der}{\der t_m} - B_m(s), \frac{\der}{\der t_n} - B_n(s)
    \right]
    = 0,
\label{cmkp:zs}
\end{gather}
or
\begin{gather}
    \left[
    \frac{\der}{\der t_m} - B^c_m(s), \frac{\der}{\der t_n} - B^c_n(s)
    \right]
    = 0,
\label{cmkp:zsc}
\end{gather}
where $B^c_n(s):= -(L(s)^n)_{\leq 0} = B_n(s) - L(s)^n$.

\subsection{Dressing operator, wave function}
\label{subsec:dressing-op/func}

Similarly to the mKP hierarchy, we can show the existence of the
dressing operator.
\begin{proposition}
\label{prop:dressing-op}
 For any solution $(L(s), P(s))_{s\in S}$ there exists an operator
 $W(s)=W(s;x,t;\der)$ of the form
\begin{gather}
    W(s;x,t;\der)
    = \bigl( w_0(s;x,t) + w_1(s;x,t) \der^{-1} + \cdots) \der^{n_s},
    \qquad
    w_0(s;x,t) \neq 0,
\label{dressing-op:diff-l}
\end{gather}
satisfying equations
\begin{gather}
L(s) W(s) = W(s) \der,\qquad  P(s) W(s) = W(s+1),
\qquad \frac{\der W(s)}{\der t_n} = B^c_n(s) W(s).
\label{lin-eq:diff-l:op}
\end{gather}
\end{proposition}
In fact, equations \eqref{cmkp:dL/dt}, \eqref{cmkp:zsc},
\eqref{cmkp:LP=PL} and \eqref{cmkp:BP=PB} are compatibility
conditions for the linear system \eqref{lin-eq:diff-l:op}.

We call $W(s;x,t;\der)$ the {\em dressing operator}.

The {\em wave function} $w(s;\lambda) = w(s;x,t;\lambda)$ is def\/ined by
\begin{gather}
 \begin{split}
    w(s;x,t;\lambda) &:=
    W(s;x,t;\der) e^{\xi(x+t,\lambda)}
\\
    &=
    \left(\sum_{j=0}^\infty w_j(s;x,t) \lambda^{-j}\right) \lambda^{n_s}
    e^{\xi(x+t,\lambda)},
 \end{split}
\label{wave-function}
\end{gather}
where $x+t:= (x+t_1, t_2, t_3, \dots)$, $\xi(x+t,\lambda) = x\lambda
+ \sum\limits_{n=1}^\infty t_n \lambda^n$. It is subject to the equations:
\begin{gather}
    L(s) w(s;\lambda) = \lambda w(s;\lambda),\nonumber
\\
    \frac{\der}{\der t_n} w(s;\lambda) = B_n(s) w(s;\lambda),\label{lin-eq:wave-func}
\\
    P(s) w(s;\lambda)= w(s+1;\lambda).\nonumber
\end{gather}

Recall that the mKP hierarchy in \cite{tak:02} is def\/ined by the
equations \eqref{cmkp:dL/dt}--\eqref{cmkp:BP=PB} for
operators $L(s)=L^\mkp(s)$, $B_n(s) = B^\mkp_n(s)$, $P(s)=P^\mkp(s)$
normalized as
\begin{gather}
    L^\mkp(s;x,t)
    = \der + u^\mkp_2(s,x,t) \der^{-1} + u^\mkp_3(s,x,t)\der^{-2} +\cdots,
\label{def:Lmkp}
\\
    B^\mkp_n(s;x,t) := (L^\mkp(s;x,t)^n)_{\geq 0},
\label{def:Bmkp}
\\
    P^\mkp(s;x,t)
    = \der^{m_s} + q_1(s,x,t) \der^{m_s-1} + \cdots + q_{m_s}(s,x,t).
\label{def:Pmkp}
\end{gather}
Its dressing operator $W(s)=W^\mkp(s)$ is normalized as
\begin{gather}
   W^\mkp(s) = (1 + w^\mkp_1(s)\der^{-1} + \cdots) \der^{n_s},
\label{Wmkp}
\end{gather}
and satisf\/ies the same linear equations \eqref{lin-eq:diff-l:op} as the
cmKP hierarchy, where
$B^c_n(s)=B^{\mkp,c}_n(s)=-(L^\mkp(s)^n)_{<0}=B^\mkp_n(s)-L^\mkp(s)^n$.

\begin{proposition}
\label{prop:cmkp=mkp+gauge}
 {\rm (i)}
 Let $(L(s), P(s))_{s\in S}$ be a solution of the cmKP hierarchy and
 $W(s)$ be the corresponding dressing operator of the form
 \eqref{dressing-op:diff-l}. Then $(L^\mkp (s), P^\mkp (s))_{s\in S}$
 defined by
\begin{gather*}
    L^\mkp (s) := w_0(s)^{-1} L(s) w_0(s), \qquad
    P^\mkp (s) := w_0(s+1)^{-1} P(s) w_0(s)
\end{gather*}
 is a solution of the mKP hierarchy and $W^\mkp (s):= w_0(s)^{-1} W(s)$
 is the corresponding dressing operator.

 {\rm (ii)}
 Conversely, if a sequence $\{(f^{(0)}(s),\dots,f^{(m_s-1)}(s))\}_{s\in
 S}$ of non-zero constant vectors and a solution of the mKP hierarchy
 $(L^\mkp(s), P^\mkp(s))_{s\in S}$ are given, there exists a unique
 function $f(s) = f(s;x,t)$ such that
\begin{gather}
    \der^k f(s;0,0) = f^{(k)}(s)
  \qquad  \text{for all}\quad s \in S, 0 \leqq k < m_s,
\label{w0:init-val}
\end{gather}
 and $(L(s):= f(s)^{-1} L^\mkp(s) f(s), P(s):= f(s+1)^{-1} P^\mkp(s)
 f(s))_{s\in S}$ is a solution of the cmKP hierarchy. $(m_{\max(S)}=\infty.)$
\end{proposition}

\begin{proof}
 (i)
 Note that the linear equations \eqref{lin-eq:diff-l:op} imply that
\begin{gather}
    \frac{\der}{\der x} w_0(s;x,t)= - u_1(s;x,t) w_0(s;x,t),
\label{dw0/dx}
\\
    w_0(s;x,t) p_0(s;x,t) = w_0(s+1;x,t),
\label{w0(s)p0(s)=w0(s+1)}
\\
    \frac{\der}{\der t_n} w_0(s;x,t) = - \bigl(L(s)^n\bigr)_0 w_0(s;x,t).
\label{dw0/dtn}
\end{gather}
 Equation \eqref{dw0/dx} and \eqref{w0(s)p0(s)=w0(s+1)} mean that
 $L^\mkp(s)=w_0(s)^{-1}L(s)w_0(s)$ and $P^\mkp(s) = w_0(s+1)^{-1} P(s)
 w_0(s)$ have the required form \eqref{def:Lmkp} and
 \eqref{def:Pmkp}. It follows from equation \eqref{dw0/dtn} that
\begin{gather*}
    w_0(s)^{-1} \left(\frac{\der}{\der t_n} - B_n(s)\right) w_0(s)
    =
    \frac{\der}{\der t_n} - B^\mkp_n(s),
\end{gather*}
 where $B^\mkp_n(s)$ is def\/ined by \eqref{def:Bmkp} from $L^\mkp(s)$. It
 is easy to see that $(L^\mkp(s), P^\mkp(s))_{s\in S}$ satisf\/ies the
 system \eqref{cmkp:dL/dt}--\eqref{cmkp:BP=PB}.

 (ii) is proved in almost the same way as the fact mentioned in
 \remref{rem:P(s)}, so we prove it in \appref{app:P(s)}.
\end{proof}

It was shown in \cite{tak:02} that dressing operators of the mKP hierarchy
are parametrized by the f\/lag manifold: Let $\calV$ be an inf\/inite
dimensional linear space $\bigoplus_{\nu\in\Integer} \Complex e_\nu$
with basis $\{e_\nu\}_{\nu\in\Integer}$ and~$V^\varnothing$ be its
subspace def\/ined by $V^\varnothing = \bigoplus_{\nu \geq 0} \Complex
e_\nu$.  (Actually we have to take completion of~$\calV$, but details
are omitted.) The {\em Sato--Grassmann manifold} of charge $n$,
$\SGM^{(n)}$, is def\/ined by
\begin{gather}
    \SGM^{(n)} =
    \{ U \subset \calV \mid
    \text{ index of\ }U \to \calV/V^\varnothing \text{ is\ }n
    \}.
\label{def:SGM}
\end{gather}
The set of dressing operators of the KP hierarchy is $\SGM^{(0)}$ as is
shown in \cite{sat:81,sat-sat:82} or \cite{sat-nou:84} and
the set of dressing operators of the mKP hierarchy is the f\/lag manifold
\begin{gather}
    \Flag :=
    \{ (U_s)_{s\in S} \mid
    U_s \in \SGM^{(n_s)}, \ U_s \subset U_{s+1}\}.
\end{gather}
See Proposition 1.3 of \cite{tak:02}. Hence the set of the dressing
operators of the cmKP hierarchy is described as follows.
\begin{corollary}
\label{cor:sol<->(phi,flag)}
 The dressing operator of the cmKP hierarchy $(W(s))_{s\in S}$ is
 parametrized by $\Flag\times \prod\limits_{s\in S'} (\Complex^{m_s}\setminus
 \{0\})$.
\end{corollary}

Schematically, this space is an inf\/inite dimensional homogeneous space
$GL(\infty)/Q$, where $Q$ is a subgroup of the group $GL(\infty)$ of
invertible $\Integer\times\Integer$ matrices def\/ined as follows:
$g=(g_{ij})_{i,j\in\Integer} \in Q$ if and only if
\begin{gather}
    g_{ij}=\begin{cases}
    0, &i> n_s,   j \leqq n_s   \text{\ or\ }
        i> n_s+1, j \leqq n_s+1 \text{\ for some\ }s\in S,
    \\
    1, &i = j = n_s+1 \text{\ for some\ }s\in S.
    \end{cases}
\label{def:Q}
\end{gather}
In fact, if we consider an intermediate parabolic subgroup $P$ def\/ined
by
\begin{gather}
    g=(g_{ij})_{i,j\in\Integer} \in P \Longleftrightarrow
    g_{ij}=
    0 \qquad (i> n_s, j \leqq n_s \text{\ for some\ }s\in S),
\label{def:P}
\end{gather}
$GL(\infty)/Q$ is considered as the f\/iber bundle
\begin{gather*}
    GL(\infty)/Q \to GL(\infty)/P,
\end{gather*}
over $GL(\infty)/P = \Flag$. A point on a f\/iber $(U_s)_{s\in S}\in\Flag$
specif\/ies a series of non-zero vectors in $U_{s+1}/U_s$ ($s\in S$). ($U_{\max(S)+1}={\mathcal V}$.)

We do not go into details of inf\/inite dimensional homogeneous spaces. In
this picture it is clear that the group $GL(\infty)$ acts on the space
of solutions transitively. The action is explicitly described in terms
of the fermionic description of the $\tau$ functions. See the end of
\secref{subsec:fermion-tau}.

\subsection{Bilinear identity}
\label{subsec:bil-id}

Recall that the wave function and the adjoint wave function of the
mKP hierarchy have the form
\begin{gather}
    w^\mkp(s;x,t;\lambda) := W^\mkp(s) e^{\xi(x+t;\lambda)}
= \hat w^\mkp(s;x,t;\lambda) \lambda^{n_s} e^{\xi(x+t;\lambda)},
\nonumber\\
    \hat w^\mkp(s;x,t;\lambda) :=
    1 + w^\mkp_1(s;x,t)\lambda^{-1} + \cdots,
\label{wave-func:mkp}
\end{gather}
and
\begin{gather}
    w^{\mkp,\ast}(s;x,t;\lambda)
    := ((W(s)^\mkp)^\ast)^{-1} e^{-\xi(x+t;\lambda)}
= \hat w^{\mkp,\ast}(s;x,t;\lambda) \lambda^{-n_s}
      e^{-\xi(x+t;\lambda)},
\nonumber\\
    \hat w^{\mkp,\ast}(s;x,t;\lambda) :=
    1 + w_1^{\mkp,\ast}(s;x,t)\lambda^{-1} + \cdots,
\label{adj-wave-func:mkp}
\end{gather}
where $A^*$ for an operator $A$ denotes its formal adjoint: $x^* = x$,
$\der^* = - \der$, $(AB)^* = B^* A^*$. These functions are characterized
by the bilinear residue identity:
\begin{gather}
    \Res_{\lambda=\infty}
    w^{\mkp}(s;x,t;\lambda)\, w^{\mkp,\ast}(s';x',t';\lambda)\, d\lambda
    = 0,
\label{bil-res:mkp}
\end{gather}
for any $x$, $t$, $x'$, $t'$ and $s' \leqq s$. See \S~1 of
\cite{tak:02} for details.

The wave function of the cmKP hierarchy is also characterized by a
bilinear identity as follows:
\begin{proposition}
\label{prop:bil-res}
 {\rm (i)}
 The wave function of the cmKP hierarchy satisfies the following
 identity:
\begin{gather}
    \Res_{\lambda=\infty}
    w(s;x,t;\lambda) \tilde w(s';x',t';\lambda)\, d\lambda
    =
    \begin{cases}
    0, &s'<s, \\ 1, &s'=s.
    \end{cases}
\label{bilinear-res}
\end{gather}
 Here $\tilde w(s,t;\lambda)$ is the adjoint wave function defined by
\begin{gather}
    \tilde w(s;x,t;\lambda) :=
    \tilde{W}(s;x,t;\der) e^{-\xi(x+t,\lambda)},
\label{adj-wave-func}
\\
    \tilde{W}(s,t;\der) = - \der^{-1} (W(s,t;\der)^{-1})^*.
\label{tildeW:cmkp}
\end{gather}

 {\rm (ii)}
 Conversely, let $w(s;x,t;\lambda)$ be a function of the form
 \eqref{wave-function} and $\tilde w(s;x,t;\lambda)$ be a function of
 the form \eqref{adj-wave-func}, where the operator $\tilde W(s)$ has
 the form
\begin{gather}
    \tilde W(s;x,t;\der)
    = (\tilde w_0(s;x,t) + \tilde w_1(s;x,t) (-\der)^{-1} + \cdots)
    (-\der)^{-n_s-1}.
\label{tilde-W}
\end{gather}
 If the pair $(w(s;x,t;\lambda), \tilde w(s;x,t;\lambda))$ satisfies the
 equation \eqref{bilinear-res}, then $w(s;x,t;\lambda)$ is a wave
 function of the cmKP hierarchy and $\tilde w(s;x,t;\lambda)$ is its
 adjoint.
\end{proposition}

\begin{proof}
 (i)
 When $s'=s$, we have only to show
\begin{gather}
    \Res_{\lambda=\infty}
    \frac{\der^\alpha w}{\der t^\alpha}(s;x,t;\lambda)
    \tilde w(s;x',t;\lambda)\, d\lambda
    = \begin{cases}
    1, &\alpha = (0,0,\dots),\\
    0, &\text{otherwise}
    \end{cases}
\label{bilinear-res:s=s':coef}
\end{gather}
 for each multi-index $\alpha=(\alpha_1,\alpha_2,\dots)$.

 Let us recall DJKM's lemma (Lemma 1.1 of \cite{djkm}): For any
 operators $A(x,\der_x)$ and $B(x,\der_x)$, we have
\begin{gather}
    \Res_{\lambda=\infty}
    A(x,\der_x) e^{x\lambda} B(x',\der_{x'}) e^{-x'\lambda}
    d\lambda
    = f(x,x'),
\label{djkm-lemma}
\end{gather}
 where $f(x,x')$ is determined by
\begin{gather}
    f(x,x') \der^{-1}\delta(x-x')
    =
    (A(x,\der_x) B^*(x,\der_x))_{<0} \delta(x-x').
\label{def:f}
\end{gather}

 Since $W(s;\der) \tilde W(s;\der)^* = \der^{-1}$ by the def\/inition
 \eqref{tildeW:cmkp} of $\tilde W(s;\der)$, we have
 \eqref{bilinear-res:s=s':coef} for $\alpha=(0,0,\dots)$ thanks to
 \eqref{djkm-lemma}.

 {\samepage When $\alpha\neq (0,0,\dots)$, we can prove by induction that
\begin{gather*}
    \frac{\der^\alpha}{\der t^\alpha} w(s;x,t;\lambda)
    =
    \sum_{i \geq 1} c^{(\alpha)}_i(s;x,t) \der^i w(s;x,t;\lambda),
\end{gather*}
 where $c^{(\alpha)}_i$'s are dif\/ferential polynomials of coef\/f\/icients
 of $B_n(s)$'s. Hence the left hand side of~\eqref{bilinear-res:s=s':coef} 
 vanishes due to \eqref{djkm-lemma}
 because
\begin{gather}
    \left(\sum_{i \geq 1} c^{(\alpha)}_i(s;x,t) \der^i\right)
    W(s;x,t;\der)
    \tilde W(s;x,t;\der)^*
    =
    \sum_{i \geq 1} c^{(\alpha)}_i(s;x,t) \der^{i-1}
\label{bilinear-res:s=s':temp}
\end{gather}
 is a dif\/ferential operator.}

 When $s'<s$, the bilinear residue identity \eqref{bilinear-res} is
 equivalent to the vanishing of its Taylor coef\/f\/icients:
\begin{gather}
    \Res_{\lambda=\infty}
    \frac{\der^\alpha w}{\der t^\alpha}(s;x,t;\lambda)
    \tilde w(s';x',t;\lambda)\, d\lambda
    = 0,
\label{bilinear-res:s'<s:coef}
\end{gather}
 for each multi-index $\alpha$. When $\alpha=(0,0,\dots)$, this follows
 directly from \eqref{djkm-lemma} since
\begin{gather*}
    W(s;\der) \tilde W(s';\der)^*
    =
    W(s;\der) W(s';\der)^{-1} \der^{-1}
\\
   \phantom{W(s;\der) \tilde W(s';\der)^*}{} =
    P(s-1) P(s-2) \cdots P(s'+1) P(s') \der^{-1},
\end{gather*}
 which is a dif\/ferential operator due to \eqref{def:P(s)}. For
 $\alpha\neq (0,0,\dots)$, the proof is similar to the case $s'=s$.

 (ii)
 When $s'=s$ and $t=t'$, the bilinear residue identity
 \eqref{bilinear-res} is equivalent to
\begin{gather*}
    (W(s;x,t;\der) \tilde W(s;x,t;\der)^*)_{<0} = \der^{-1},
\end{gather*}
 which means
\begin{gather*}
     W(s;x,t;\der) \tilde W(s;x,t;\der)^* = \der^{-1},
\end{gather*}
because $W(s;x,t;\der) \tilde W(s;x,t;\der)^*$ is of order $-1$.
Hence
\begin{gather}
    \tilde W(s;x,t;\der) = -\der^{-1} (W(s;x,t;\der)^{-1})^*.
\label{tildeW=W-1*}
\end{gather}

 Putting $s'=s-1$ and $t=t'$, we have
\begin{gather*}
    (W(s;x,t;\der) \tilde W(s-1;x,t;\der)^*)_{<0} = 0.
\end{gather*}
 from \eqref{bilinear-res}. This means
\begin{gather*}
    W(s;x,t;\der) W(s-1;x,t;\der)^{-1} \der^{-1}
    = (\text{dif\/ferential operator}),
\end{gather*}
 by \eqref{tildeW=W-1*}. Hence we obtain
\begin{gather}
    P(s-1;\der) := W(s,t;\der) W(s-1,t;\der)^{-1}
\nonumber\\
 \phantom{P(s-1;\der) }{}   = (\text{dif\/ferential operator of order\ }m_{s-1}
       \text{\ divisible by\ }\der).
\label{WW=P}
\end{gather}
 Finally, put $s'=s$ and dif\/ferentiate \eqref{bilinear-res} with respect
 to $t_n$. Then we have
\begin{gather*}
    \Res_{\lambda=\infty} \frac{\der}{\der t_n} w(s,t;\lambda)
    \tilde w(s,t;\lambda) d\lambda = 0,
\end{gather*}
 namely,
\begin{gather*}
    \left(
    \left(\frac{\der W(s)}{\der t_n} + W(s) \der^n\right)
    \tilde W(s)^*
    \right)_{<0} = 0.
\end{gather*}
 Using \eqref{tildeW=W-1*}, we can rewrite this equation as
\begin{gather*}
    \frac{\der W(s)}{\der t_n} W(s)^{-1}
    =
    - (W(s) \der^n W(s)^{-1})_{\leq 0}.
\end{gather*}
 Thus we have recovered the linear equations \eqref{lin-eq:diff-l:op}.
\end{proof}

\begin{corollary}
\label{cor:cmkp->mkp}
 The function $w^\mkp(s;x,t;\lambda):=w(s;x,t;\lambda)/w_0(s;x,t)$ is a
 wave function of the mKP hierarchy. Its adjoint wave function is
\[
    w^{\mkp,*}(s;x,t;\lambda) 
    :=
    - w_0(s;x,t) \der(\tilde w(s;x,t;\lambda)).
\] 
\end{corollary}

\begin{proof}
 This can be directly deduced from \propref{prop:cmkp=mkp+gauge}.
 Alternatively we derive it from \propref{prop:bil-res} here. Functions
 $w(s;x,t;\lambda)/w_0(s;x,t)$ and $-w_0(s;x,t)\der(\tilde
 w(s;x,t;\lambda))$ are expanded with respect to $\lambda$ as in
 \eqref{wave-func:mkp} and in \eqref{adj-wave-func:mkp}
 respectively. Hence dif\/ferentiating \eqref{bilinear-res} with respect
 to~$x'$, we obtain the bilinear residue identity \eqref{bil-res:mkp}
 for the mKP hierarchy.
\end{proof}

\subsection[$\tau$ function]{$\boldsymbol{\tau}$ function}
\label{subsec:tau}

In this subsection we prove that the wave functions of the cmKP
hierarchy are ratios of the $\tau$~functions. In contrast to the (m)KP
hierarchy, we need two series of $\tau$ functions to express the wave
function, unless $m_s=1$.

\begin{theorem}
\label{thm:tau}
{\rm (i)}
 Let $w(s;x,t;\lambda)$ be a wave function of the cmKP hierarchy and
 $\tilde w(s;x,t;\lambda)$ be its adjoint. Then there exist functions
 $\tau_0(s;t)$ and $\tau_1(s;t)$ such that
\begin{gather}
    w(s;x,t;\lambda) :=
    \frac{\tau_0(s;x+t-[\lambda^{-1}])}
         {\tau_1(s;x+t)} \lambda^{n_s} e^{\xi(x+t;\lambda)},
\label{w=tau/tau}
\\
    \tilde w(s;x,t;\lambda) :=
    \frac{\tau_1(s;x+t+[\lambda^{-1}])}
         {\tau_0(s;x+t)} \lambda^{-n_s-1} e^{-\xi(x+t;\lambda)},
\label{tildew=tau/tau}
\end{gather}
 where $[\lambda^{-1}] := (\lambda^{-1}, \lambda^{-2}/2, \lambda^{-3}/3,
 \dots)$. The {\em $\tau$ functions} $\tau_0(s;t)$ and $\tau_1(s;t)$ are
 determined only up to multiplication by an arbitrary function of $s$.

{\rm (ii)}
 If\/ $m_s=n_{s+1}-n_s=1$, we can choose $\tau$ functions so that
 $\tau_1(s;t) = \tau_0(s+1;t)$.

{\rm (iii)}
 The $\tau$ functions are characterized by the following bilinear
 residue identity:
\begin{gather}
    \Res_{\lambda=\infty}
    \tau_0(s;t-[\lambda^{-1}]) \tau_1(s';t'+[\lambda^{-1}])
    e^{\xi(t,\lambda) - \xi(t',\lambda)} \lambda^{n_s-n_{s'}-1}
    \, d\lambda
\nonumber\\
 \qquad\qquad{}   =
    \begin{cases}
    0, & s'<s, \\ \tau_1(s;t)\, \tau_0(s;t'), &s'=s.
    \end{cases}
\label{bilinear-res:tau}
\end{gather}
\end{theorem}

\begin{proof}
 (i)
 Let us denote the non-trivial parts of the wave function
 \eqref{wave-function} and the adjoint wave function
 \eqref{adj-wave-func} as follows:
\begin{gather*}
    \hat{w}(s;x,t;\lambda) :=
    w_0(s;x,t) + w_1(s;x,t)\lambda^{-1} + w_2(s;x,t)\lambda^{-2} + \cdots,
\\
    \hat{\tilde w}(s;x,t;\lambda) :=
    \tilde w_0(s;x,t) + \tilde w_1(s;x,t)\lambda^{-1} +
    \tilde w_2(s;x,t)\lambda^{-2} + \cdots.
\end{gather*}
 Namely, 
 \begin{gather*}
    w(s;x,t;\lambda)
    =
    \hat{w}(s;x,t;\lambda) \lambda^{n_s} e^{\xi(x+t;\lambda)},
\\
    \tilde w(s;x,t;\lambda)
    =
    \hat{\tilde w}(s;x,t;\lambda) \lambda^{-n_s-1} e^{-\xi(x+t;\lambda)}.
 \end{gather*}
 Putting $s=s'$, $x=x'=0$, replacing $t_n$ by $t_n+\zeta^{-n}/n$, $t_n'$
 by $t_n$ in the bilinear identity \eqref{bilinear-res}, we have
\begin{gather}
    1=\Res_{\lambda=\infty}
    \left(
     \hat{w}        (s;t+[\zeta^{-1}];\lambda)
     \hat{\tilde w}(s;t             ;\lambda)
     \frac{\lambda^{-1}}{1-\lambda\zeta^{-1}}
    \right)\, d\lambda
\nonumber\\
\phantom{1}{}    =\hat{w}(s;t+[\zeta^{-1}];\zeta)\hat{\tilde w}(s;t;\zeta).
\label{hatw(s)hattildew(s)=1}
\end{gather}
 In the limit $\zeta^{-1}\rightarrow 0$ we have
\begin{gather}
    w_0(s;t)\tilde w_0(s;t)=1.
\label{w0w0*=1}
\end{gather}
 Since $w_0(s;t)^{-1}w(s;t;\lambda)$ is a wave function of the mKP
 hierarchy (cf.\ \corref{cor:cmkp->mkp}), there exists a tau function
 $\tau_0(s;t)$ such that
\begin{gather}
    \frac{\hat{w}(s;t;\lambda)}{w_0(s,t)}
    =\frac{\tau_0(s;t-[\lambda^{-1}])}{\tau_0(s;t)}.
\label{hatw/w0=tau/tau}
\end{gather}
 Def\/ine the function $\tau_1(s;t)$ by{\samepage
\begin{gather}
    \tau_1(s;t) := \frac{\tau_0(s;t)}{w_0(s;t)}.
\label{tau1=tau0/w0}
\end{gather}
 Equation \eqref{w=tau/tau} follows from \eqref{hatw/w0=tau/tau} and
 \eqref{tau1=tau0/w0}. Equation \eqref{tildew=tau/tau} follows from
 \eqref{hatw(s)hattildew(s)=1}.}

(ii)
 From \eqref{tildew=tau/tau} and \eqref{hatw(s)hattildew(s)=1}, we can
 see that dependence of $\tau_0(s;t)$ on $t_n$ ($n \geq 1$) is
 determined uniquely by the equation
\begin{gather}
    \frac{\der\log\tau_0(s;t)}{\der t_n}= A_n(s;t),
\label{tau4}
\end{gather}
 where
\begin{gather*}
    A_n(s;t) = -\Res_{\lambda=\infty}
    \left(\lambda^n
     \left(
      \sum_{j=1}^{\infty}\lambda^{-j-1} \frac{\der}{\der t_j}
      +                                 \frac{\der}{\der\lambda}
     \right)\log \hat{\tilde w}(s;t;\lambda)
    \right)\, d\lambda.
\end{gather*}
 See \S~1.6 of \cite{djkm} for detailed arguments.

 When $n_{s+1}=n_s+1$, by putting $s'=s-1$ and replacing $t_n$ by
 $t_n+\zeta^{-n}/n$, $t_n'$ by $t_n$ in the bilinear identity
 \eqref{bilinear-res}, we have
\begin{gather*}
    0=\Res_{\lambda=\infty}\left(
      \hat{w}       (s  ;t+[\zeta^{-1}];\lambda)
      \hat{\tilde w}(s-1;t             ;\lambda)
      \frac{1}{1-\lambda\zeta^{-1}}\right)\, d\lambda
\\
 \phantom{0}{}    =
     \hat{w}       (s  ;t+[\zeta^{-1}];\zeta)
     \hat{\tilde w}(s-1;t             ;\zeta)
     - w_0(s;t+[\zeta^{-1}]) \tilde w_0(s-1;t).
\end{gather*}
 Using \eqref{hatw(s)hattildew(s)=1} and \eqref{w0w0*=1}, and putting
 $\zeta=\lambda$, we obtain
\begin{gather*}
    \frac{\hat{\tilde w}(s-1;t;\lambda)}{\hat{\tilde w}(s;t;\lambda)}
    =
    \frac{w_0(s;t+[\lambda^{-1}])}{w_0(s-1;t)}
\end{gather*}
 Therefore, we have from def\/inition \eqref{tau4},
\begin{gather*}
    \frac{\der\log\tau_0(s;t)}{\der t_n}
    -
    \frac{\der\log\tau_0(s-1;t)}{\der t_n}
    =
    -\frac{\der\log w_0(s-1;t)}{\der t_n}.
\end{gather*}
 Hence, we can f\/ix the dependence of $\tau_0(s;t)$ on $s$ by
\begin{gather*}
    \log \tau_0(s;t)-\log \tau_0(s-1;t)=-\log w_0(s-1;t).
\end{gather*}
 Comparing this with the def\/inition \eqref{tau1=tau0/w0}, we f\/ind that
 $\tau_1(s;t)=\tau_0(s+1;t)$, as desired.

 Statement (iii) is a direct consequence of (i) and the bilinear residue
 identity for the wave functions, \eqref{bilinear-res}.
\end{proof}

\subsection[Construction of $\tau$ function]{Construction of $\boldsymbol{\tau}$ function}
\label{subsec:fermion-tau}

In this subsection we construct $\tau$ functions of the cmKP hierarchy
in terms of the free fermions or, in other words, the Clif\/ford algebra
as in the case of the KP hierarchy \cite{djkm} or of the Toda lattice
hierarchy \cite{tak:91a,tak:91b}.

Let $\psi_n$ and $\psi^*_n$ ($n\in\Integer$) be free fermion
operators, i.e., generators of a Clif\/ford algebra $\calA$ which
satisfy the canonical anti-commutation relations:
\begin{gather}
    [\psi_m, \psi_n]_+ = [\psi^*_m, \psi^*_n]_+ = 0, \qquad
    [\psi_m, \psi^*_n]_+ = \delta_{mn},
\label{car}
\end{gather}
where $[A,B]_+ := AB + BA$. The Fock space $\calF$ and the dual Fock
space $\calF^\vee$ are generated by the vacuum vector
$|\mathrm{vac}\rangle$ and its dual $\langle \mathrm{vac}|$ over
$\calA$ respectively. $\calF$ and $\calF^\vee$ contain states of
charge $k$, $|k\rangle$ and $\langle k|$ respectively, which are
characterized by
\begin{align}
    &\psi_n |k\rangle =
    \begin{cases}
      0,            &n < k, \\
      |k+1 \rangle, &n = k,
    \end{cases}
    &
    &\psi^*_m |k\rangle =
    \begin{cases}
      0,            &m \geqq k , \\
      |k-1 \rangle, &m =     k-1,
    \end{cases}&
\label{|k>}
\\
    &\langle k| \psi_n =
    \begin{cases}
      0 ,            &n \geqq k , \\
      \langle k-1 |, &n =     k-1,
    \end{cases}
    &
    &\langle k| \psi^*_m=
    \begin{cases}
      0 ,            &m < k, \\
      \langle k+1 |, &m = k.
    \end{cases}&
\label{<k|}
\end{align}
In fact, $|\mathrm{vac}\rangle = |0\rangle$ and
$\langle\mathrm{vac}|=\langle0|$. The pairing of $\calF$ and
$\calF^\vee$ is naturally def\/ined by $\langle k | k \rangle = 1$.

We def\/ine the operators $J(t)$, $\psi(\lambda)$ and
$\psi^*(\lambda)$ as follows:
\begin{gather}
    J(t) :=
    \sum_{n=1}^\infty t_n \sum_{k\in\Integer} \psi_k \psi^*_{k+n},
\label{def:J}
\\
    \psi(\lambda) := \sum_{n\in\Integer} \psi_n \lambda^n, \qquad
    \psi^*(\lambda) := \sum_{n\in\Integer} \psi^*_n \lambda^{-n}.
\label{def:psi(l),psi*(l)}
\end{gather}
We quote important formulae from \S~2.6 of \cite{djkm}:
\begin{gather}
    \langle m | e^{J(t)} \psi(\lambda)
    =
    \lambda^{m-1} e^{\xi(t,\lambda)} \langle m-1 |
    e^{J(t-[\lambda^{-1}])},
\label{<m|eJpsi}
\\
    \langle m | e^{J(t)} \psi^*(\lambda)
    =
    \lambda^{-m} e^{-\xi(t,\lambda)} \langle m+1 |
    e^{J(t+[\lambda^{-1}])}.
\label{<m|eJpsi*}
\end{gather}
The bilinear identity comes from the following intertwining relation
\cite[\S~2.1]{djkm}:
\begin{gather}
    \sum_{n\in\Integer} \psi_n g \tensor \psi^*_n g
    =
    \sum_{n\in\Integer} g \psi_n \tensor g \psi^*_n,
\label{intertwining}
\end{gather}
where $g$ is an arbitrary element of the Clif\/ford group generated by
$\psi_n$'s and $\psi^*_n$'s.

Putting this equation between $\langle m+1| e^{J(t)} \tensor \langle
m'-1| e^{J(t')}$ and $|m \rangle \tensor |m' \rangle$, we have
\begin{gather*}
\Res_{\lambda=\infty}
    \langle m+1  | e^{J(t )} \psi  (\lambda) g |m \rangle
    \langle m'-1 | e^{J(t')} \psi^*(\lambda) g |m'\rangle
    \frac{d\lambda}{\lambda}
\\
\qquad{} =\sum_{n \in \Integer}
    \langle m+1  | e^{J(t )} \psi_n   g |m \rangle
    \langle m'-1 | e^{J(t')} \psi^*_n g |m'\rangle
\\
  \qquad{}  =\sum_{n \in \Integer}
    \langle m+1  | e^{J(t )} g \psi_n   |m \rangle
    \langle m'-1 | e^{J(t')} g \psi^*_n |m'\rangle.
\end{gather*}
Thus we obtain
\begin{gather}
\Res_{\lambda=\infty}
    \langle m+1  | e^{J(t )} \psi  (\lambda) g |m \rangle
    \langle m'-1 | e^{J(t')} \psi^*(\lambda) g |m'\rangle
    \frac{d\lambda}{\lambda}
\nonumber\\
 \qquad{}   = \begin{cases}
    0, & m' \leqq m,
\\
    \langle m+1  | e^{J(t )} g |m+1 \rangle
    \langle m    | e^{J(t')} g |m   \rangle,
    & m'=m+1
    \end{cases}
\label{res-id:0}
\end{gather}
due to \eqref{|k>}.

For a Clif\/ford group element $g$ we def\/ine $\tau$ functions by
\begin{gather}
    \tau_0(s;t) := \langle n_s   | e^{J(t)} g | n_s   \rangle, \qquad
    \tau_1(s;t) := \langle n_s+1 | e^{J(t)} g | n_s+1 \rangle.
\label{def:tau0,tau1}
\end{gather}
The bilinear residue identity \eqref{bilinear-res:tau} holds because of
\eqref{<m|eJpsi}, \eqref{<m|eJpsi*} and \eqref{res-id:0}. Namely, we
have constructed a pair of $\tau$ functions of the cmKP hierarchy for
each $g$ in the Clif\/ford group.

The action of $GL(\infty)$ mentioned at the end of
\secref{subsec:dressing-op/func} is realized as the action of the Clif\/ford
group, $g \mapsto g'g$ ($g'\in GL(\infty)$) in
\eqref{def:tau0,tau1}. Hence the above construction exhausts all the
solutions of the cmKP hierarchy.

The vertex operator description of the $gl(\infty)$ symmetry is the same
as that for the KP hierarchy in \cite{djkm}.


\section[Dispersionless modified KP hierarchy]{Dispersionless modif\/ied KP hierarchy}
\label{sec:dcmkp}
\subsection[Definition of the dcmKP hierarchy]{Def\/inition of the dcmKP hierarchy}
Let $N$ be a positive integer. When $n_s=Ns$, we can introduce the
parameter $\hbar$ into the cmKP hierarchy and take the dispersionless
limit, as is done for the mKP hierarchy in \cite{tak:02}. Now for the
dcmKP hierarchy, the independent variables are the continuous variables
$s,x$ and $t=(t_1, t_2, \ldots).$ The dependent variables are
encapsulated in $\mL(k;s,t)$ and $\mP(k;s,t)$, which are respectively
formal power series and polynomial of $k$ having the following form:
\begin{gather}
\mL(s)=\mL(k;s,t)=k+u_1(s,t)+u_2(s,t)k^{-1}+\cdots=\sum_{n=0}^{\infty}u_n(s,t)k^{1-n} \label{definemL}, \\
 \mP(s)=\mP(k;s,t)= p_0(s,t)k^{N} + \cdots +
p_{N-1}(s,t)k=\sum_{n=0}^{N-1} p_n(s,t)k^{N-n}.\label{definemP}
\end{gather}
Here $u_0=1$, $p_0\neq 0$. We do not write the dependence on $x$
explicitly for the reason we are going to see later.

The $N$-dcmKP hierarchy is the following system of equations:
\begin{gather}
\frac{\pa\mL}{\pa t_n} = \{\B_n, \mL\}, \qquad \B_n:
=(\mL^n)_{>0},\label{equation1}\\
\frac{\pa \mL}{\pa s} = \{ \log \mP, \mL\}
,\label{equation2}\\
\frac{\pa \log \mP}{\pa t_n}=\frac{\pa \B_n }{\pa s} - \{\log\mP,
\B_n\},\label{equation3}
\end{gather}
where now the projection $(\cdot )_{>0}$ is with respect to $k$,
$\log\mP$ is formally understood as
\begin{gather*}
\log\mP=\log p_0+ \log k^N +\sum_{n=1}^{\infty} \mathfrak{p}_n
k^{-n},
\end{gather*}
and $\{\cdot, \cdot\}$ is the Poisson bracket
\begin{gather*}
\{ f(k,x),g(k,x)\}=\frac{\pa f}{\pa k}\frac{\pa g}{\pa x}-\frac{\pa
f}{\pa x}\frac{\pa g}{\pa k}.
\end{gather*}
 As usual, the $n=1$ case of
equations \eqref{equation1} and \eqref{equation3} implies that the
dependence on $x$ and $t_1$ appears in the combination $x+t_1$. As a
result, we usually omit $x$ and identify $t_1$ with $x$.

By the standard argument, equation \eqref{equation1} is equivalent
to the Zakharov--Shabat (or zero-curvature)  equations
\begin{gather}
  \frac{\pa \B_n}{\pa t_m}-\frac{\pa \B_m}{\pa t_n} +\{ \B_n, \B_m\}=0,\qquad 
  \text{or}\qquad  
  \frac{\pa \B_n^c}{\pa t_m}-\frac{\pa \B_m^c}{\pa t_n}
  +\{\B_n^c, \B_m^c\} = 0,
\label{Z-S}
\end{gather}
where $\B_n^c =-(\mL^n)_{\leq 0} = \B_n -\mL^n$.

\begin{remark}
\label{remark1}
 As in the case of the cmKP hierarchy (cf.\ \remref{rem:P(s)}), even
 when we start from polynomial $\mP$ with a constant term, $\mP(s) =
 p_0(s)k^N + \cdots + p_N(s)$, we can gauge away $p_N(s)$. See
 \appref{app:P(s)}. 

 In general, we can let $\mP$ be a power series with leading term
 $p_0k^N$, and have inf\/initely many negative power terms, i.e.,
 $\mP=\sum\limits_{n=0}^{\infty} p_n k^{N-n}$. In particular, if $\mP=\mL^N$,
 then equation \eqref{equation2} says that there is no dependence on
 $s$, and \eqref{equation3} is equivalent to \eqref{equation1}. This is
 the dmKP hierarchy considered by Kupershmidt \cite{Kuper}, Chang and Tu
 \cite{CT}.
\end{remark}

\begin{remark}
Suppose $(\mL, \mP)$ is a solution of N-dcmKP hierarchy
\eqref{equation1}--\eqref{equation3} and
$Q(\zeta)=a_m\zeta^m+\cdots+a_0$ is a polynomial with coef\/f\/icients
$a_0, \ldots, a_m$ independent of $s$, $x$ and $t$. If
$Q(\mL(k))\mP(k)$ is a polynomial of $k$ without constant
term\footnote{We do not need to impose this condition when we
consider the generalized cmKP hierarchy as in Remark
\ref{remark1}.}, then it is easy to see that $(\mL, \mP Q(\mL))$ is
a solution of $(N+m)$-dcmKP hierarchy. We say that this solution is
equivalent to the solution $(\mL,\mP)$.
\end{remark}

\subsection{Dressing operator} As in \cite{TT1}, we can show the
existence of a dressing operator.

\begin{proposition}\label{varphi}
For any solution $(\mL(s),\mP(s))$ of the dcmKP hierarchy, there
exists an operator $\exp \ad  \phi(s)$ that satisfies
\begin{gather}
\mL=(\exp \ad  \phi) k,\nonumber\\
\nabla_{t_n, \phi} \phi = \B_n^c, \hspace{1cm} \nabla_{s, \phi}
\phi = \log \mP - \log \mL^N,\label{dressing}
\end{gather}where $\phi(s)$ is a power series of the form
$\phi(s)=\sum\limits_{n=0}^{\infty}\phi_n(s;t)k^{-n}$, $(\ad f)g =\{f,g\}$
and
\begin{gather*}
\nabla_{u,\psi}\phi
=\sum_{m=0}^{\infty}\frac{1}{(m+1)!}(\ad\psi)^m\frac{\pa\phi}{\pa u}
\end{gather*}for series $\psi$, $\phi$ and variable $u$.
\end{proposition}

Comparing the coef\/f\/icients of $k^0$ on both sides of equations in
\eqref{dressing}, we have

\begin{corollary}\label{phi}
The function $\phi_0(s,t)$ satisfies
\begin{gather*}
\frac{\pa\phi_0}{\pa t_n} = -(\mL^n)_0, \qquad \frac{\pa
\phi_0}{\pa s} = \log p_0.
\end{gather*}
\end{corollary}

\subsection[Orlov-Schulman function]{Orlov--Schulman function}
Using the dressing operator $\phi$, we can construct the
Orlov--Schulman function $\M$ by
\begin{gather}
\M = e^{\ad \phi} \left(\sum_{n=1}^{\infty} n t_n k^{n-1} + x +
Nsk^{-1} \right)\nonumber\\
\phantom{\M}{}= \sum_{n=1}^{\infty} n t_n \mL^{n-1} + x + \frac{Ns}{\mL} +
\sum_{n=1}^{\infty} v_n \mL^{-n-1}\label{M},
\end{gather}
where $v_n$ are functions of $s$, $t$. $\M$ has the property that it
forms a canonical pair with $\mL$, namely
\begin{gather}\label{equation5}
\{ \mL, \M\}=1.
\end{gather} Using Lemma A.1
in Appendix A of \cite{TT1} and Proposition
\ref{varphi}, we  f\/ind that
\begin{gather}\label{equation4}
\frac{\pa \M}{\pa t_n} = \{ \B_n , \M\}, \qquad \frac{\pa
\M}{\pa s} = \{ \log \mP, \M\}.
\end{gather}

As in \cite{TT2}, we can show by using the equations
\eqref{equation1}--\eqref{equation3},
\eqref{equation4}, \eqref{equation5} and Corollary~\ref{phi} that
the expansion of $\B_n$ and $\log\mP$ with respect to $\mL$ can be
expressed through the functions~$v_n$. More precisely,
\begin{proposition}\label{re1}We have the following relations:
\begin{gather}
\B_n=\mL^n +\frac{\pa \phi_0}{\pa t_n}
-\sum_{m=1}^{\infty}\frac{1}{m}\frac{\pa v_m}{\pa t_n} \mL^{-m},\label{B}\\
\log\mP = \log \mL^N +\frac{\pa\phi_0}{\pa
s}-\sum_{m=1}^{\infty}\frac{1}{m}\frac{\pa v_m}{\pa s}
\mL^{-m}\label{P}.
\end{gather}
\end{proposition}

\subsection[Fundamental two form and $S$ function]{Fundamental two form and $\boldsymbol{S}$ function}
\label{subsec:2-form-and-S}
The fundamental two form $\omega$ is def\/ined by
\begin{gather*}
\omega = dk \wedge dx + \sum_{n=1}^{\infty} d\B_n \wedge dt_n + d
\log \mP   \wedge ds.
\end{gather*}
The exterior derivative $d$ is taken with respect to the independent
variables $k$, $x$, $s$ and $t $. From def\/inition, $\omega$ is closed
\[
d\omega =0,
\] and it follows from the zero-curvature equation
\eqref{Z-S} and equation \eqref{equation3} that
\[ \omega \wedge
\omega =0.
\] $(\mL, \M)$ is a pair of functions that play the role
of Darboux coordinates. Namely
\[
d\mL \wedge d\M = \omega.
\]
In fact, we can prove as Proposition 2 in \cite{TT2} that
\begin{proposition}\label{omega}
The system of equations \eqref{equation1}--\eqref{equation3}, \eqref{equation4} and \eqref{equation5} are
equivalent to
\begin{gather}\label{twoform}
d\mL \wedge d\M =dk \wedge dx + \sum_{n=1}^{\infty} d\B_n \wedge
dt_n + d \log \mP  \wedge ds.
\end{gather}
\end{proposition}

This formula implies that there exists a function $S(\mL;s,t)$ such
that
\begin{gather*}
dS = \M d\mL + k dx +\sum_{n=1}^{\infty} \B_n dt_n + \log\mP ds,
\end{gather*}
or equivalently,
\begin{gather*}
  \frac{\pa S}{\pa \mL}=\M, \qquad
  \frac{\pa S}{\pa x} = \frac{\pa S}{\pa t_1}= k, \qquad
  \frac{\pa S}{\pa t_n} =\B_n, \qquad
  \frac{\pa S}{\pa s} =\log\mP.
\end{gather*}
From the formula \eqref{M} and Proposition \ref{re1}, it is easy to
see that
\begin{proposition}
\label{prop:S}
The $S$ function is given explicitly by
\begin{gather*}
S= \sum_{n=1}^{\infty} t_n \mL^n +x\mL + s\log \mL^N
-\sum_{n=1}^{\infty} \frac{v_n}{n}\mL^{-n} +\phi_0.
\end{gather*}
\end{proposition}

\subsection{Tau function}
We introduce the power series $\mathsf{k}(z;s,t)$ as the
(formal) inverse of $\mL(k;s,t)$ with respect to $k$, i.e.\
$\mL(\mathsf{k}(z;s,t);s,t)=z$ and $\mathsf{k}(\mL(k;s,t);s,t)=k$.
Def\/ine the Grunsky coef\/f\/icients $b_{mn}$, $m,n>0$ and $b_{n0}=b_{0n}$,
$n>0$ of $\mathsf{k}(z)=\mathsf{k}(z;s,t)$ (cf.~\cite{Duren, Pom,
Teo2}) by the expansions
\begin{gather}
\log\frac{\mathsf{k}(z_1)-\mathsf{k}(z_2)}{z_1-z_2} =
-\sum_{m=1}^{\infty}b_{mn}
z_1^{-m} z_2^{-n}\label{Gr1},\\
\log \frac{\mathsf{k}(z)}{z} = -\sum_{n=1}^{\infty} b_{n0}
z^{-n}\label{Gr2}.
\end{gather}
Obviously, $b_{mn}$ are symmetric. In terms of the
Grunsky coef\/f\/icients, we have (cf.~\cite{Duren, Pom, Teo2})
\begin{gather}
  (\mL^n)_0=nb_{n,0}, \qquad 
  \B_n = \mL^n -nb_{n,0} +n\sum_{m=1}^{\infty}b_{nm} \mL^{-m}. 
\label{re3}
\end{gather}
Comparing with \eqref{B}, we f\/ind that
\begin{gather}\label{re2}
\frac{\pa v_m}{\pa t_n} = -nm b_{nm},\qquad \frac{\pa
\phi_0}{\pa t_n}=-n b_{n,0} .
\end{gather} Therefore by the symmetry of Grunsky coef\/f\/icients, the
f\/irst equation gives
\begin{gather}\label{sym1}
\frac{\pa v_m}{\pa t_n}=\frac{\pa v_n}{\pa t_m}.
\end{gather} Consequently, we have
\begin{proposition}\label{tau2}
There exists a tau function $\tau_{\dcmKP}(s;t)$, determined up
to a function of $s$, such that
\begin{gather}\label{tau} \frac{\pa
 \log\tau_{\dcmKP}}{\pa t_n}=v_n.
\end{gather}
\end{proposition}
Def\/ine $\F=\log\tau_{\dcmKP}$. It is called the free energy. Using
equations \eqref{tau} and \eqref{re2}, we can rewrite the equations
\eqref{Gr1}, \eqref{Gr2} as
\begin{gather}
z_1\exp\left(\sum_{n=1}^{\infty}\frac{1}{n}\frac{\pa\phi_0}{\pa
t_n}z_1^{-n}\right)-z_2\exp\left(\sum_{n=1}^{\infty}\frac{1}{n}\frac{\pa\phi_0}{\pa
t_n}z_2^{-n}\right)\nonumber\\
\qquad{} =(z_1-z_2)\exp\left(\sum_{m=1}^{\infty}\sum_{n=1}^{\infty}\frac{1}{mn}\frac{\pa^2\F}{\pa
t_m\pa t_n} z_1^{-m} z_2^{-n}\right).\label{dHirota}
\end{gather}
Comparing the coef\/f\/icients of $z_2^{0}$ on both sides, we
have
\begin{gather}\label{new}
z\exp\left(\sum_{n=1}^{\infty}\frac{1}{n}\frac{\pa\phi_0}{\pa
t_n}z^{-n}\right)=z+\frac{\pa\phi_0}{\pa
t_1}-\sum_{n=1}^{\infty}\frac{1}{n}\frac{\pa^2\F}{\pa t_n \pa t_1}
z^{-n}.
\end{gather}
 On the other hand, we can formulate a partial converse of
 Proposition \ref{tau2} as:
\begin{proposition}\label{prop2}
If $\tau_{\dcmKP}(s,t)$ and $\phi_0(s,t)$ are functions that satisfy
the equation \eqref{dHirota}, then the pair of functions
$(\mL,\mP)$, where $\mL(k)=\mL(k;s,t)$ is defined by taking the
inverse of the formal power series
\begin{gather*}
\mathsf{k}(z)=\mathsf{k}(z;s,t)
=z\exp\left(\sum_{n=1}^{\infty}\frac{1}{n}\frac{\pa\phi_0}{\pa
t_n}z^{-n}\right)
\end{gather*}and  $\mP(k)=\mP(k;s,t)$ is defined so that its
composition with $\mathsf{k}(z)$ is given by
\begin{gather}
\mP(\mathsf{k}(z))=\mP(\mathsf{k}(z;s,t);s,t)=
  \exp\left(\frac{\pa\phi_0}{\pa s}\right) z^N
\exp\left(-\sum_{n=1}^{\infty}\frac{1}{n}\frac{\pa^2\log\tau_{\dcmKP}}{\pa
s\pa t_n} z^{-n}\right),
\end{gather}
satisfy the dcmKP hierarchy
\eqref{equation1}--\eqref{equation3}, in the generalized sense as
Remark \ref{remark1}.
\end{proposition}
\begin{proof}
From \eqref{new}, we have \begin{gather*}
\mathsf{k}(z)=z+\frac{\pa\phi_0}{\pa
t_1}-\sum_{n=1}^{\infty}\frac{1}{n}\frac{\pa^2\F}{\pa t_n \pa t_1}
z^{-n}.
\end{gather*} Therefore, it   follows immediately from the def\/inition of
$\log\mP$ that
\begin{gather*}
\left.\frac{\pa \log\mP}{\pa t_1}\right|_{\mL
\;\text{f\/ixed}}=\left.\frac{\pa \mathsf{k}(\mL)}{\pa s}\right|_{\mL
\;\text{f\/ixed}}.
\end{gather*}
As in the proof of Proposition 3.2 in \cite{Teo2}, this
implies equation \eqref{equation2}.

On the other hand, let $b_{mn}$, $m,n\geq 0$ be the Grunsky
coef\/f\/icients of $\mathsf{k}(z)$ def\/ined as in \eqref{Gr1},
\eqref{Gr2}. Comparing the equations \eqref{dHirota}  with
\eqref{Gr1}, \eqref{Gr2}, we f\/ind from \eqref{re3} that
\begin{gather*}
\B_n = \mL^n +\frac{\pa \phi_0}{\pa t_n}
-\sum_{n=1}^{\infty}\frac{1}{m}\frac{\pa^2\log\tau_{\dcmKP}}{\pa
t_m\pa t_n}\mL^{-m}.
\end{gather*}Therefore,
\begin{gather}\label{re5}
\left.\frac{\pa \B_n}{\pa t_1}\right|_{\mL
\;\text{f\/ixed}}=\left.\frac{\pa \mathsf{k}(\mL)}{\pa
t_n}\right|_{\mL \;\text{f\/ixed}},\qquad \left.\frac{\pa
\B_n}{\pa s}\right|_{\mL \;\text{f\/ixed}}=\left.\frac{\pa
\log\mP}{\pa t_n}\right|_{\mL \;\text{f\/ixed}}.
\end{gather} The f\/irst equation
implies equation \eqref{equation1}. On the other hand, by using the
second equation in~\eqref{re5} and equations \eqref{equation1} and
\eqref{equation2}, we have
\begin{gather*}
\frac{\pa \log\mP}{\pa t_n}-\frac{\pa \B_n}{\pa s} +\{\log\mP,
\B_n\}\\
\qquad{}=\left.\frac{\pa \log\mP}{\pa t_n}\right|_{\mL
\;\text{f\/ixed}}+\frac{\pa\log\mP}{\pa\mL}\frac{\pa\mL}{\pa t_n}
-\left.\frac{\pa \B_n}{\pa s}\right|_{\mL
\;\text{f\/ixed}}-\frac{\pa\B_n}{\pa\mL}\frac{\pa\mL}{\pa s}\\
\qquad{}+\frac{\pa \log\mP}{\pa\mL}\frac{\pa\mL}{\pa k} \left.\frac{\pa
\B_n}{\pa t_1}\right|_{\mL\;\text{f\/ixed}}  - \left.\frac{\pa
\log\mP}{\pa t_1}\right|_{\mL\;\text{f\/ixed}}
\frac{\pa\B_n}{\pa\mL}\frac{\pa\mL}{\pa k}=0.
\end{gather*}This gives equation \eqref{equation3}.
\end{proof}

\subsection{Relation with the dmKP hierarchy}
The dmKP hierarchy in \cite{tak:02} is def\/ined by the system of
equations
\begin{gather}
\frac{\pa\mL^{\dmKP}}{\pa t_n} = \{\B_n^{\dmKP}, \mL^{\dmKP}\},
 \label{equ1}\\
\frac{\pa \mL^{\dmKP}}{\pa s} = \{ \log \mP^{\dmKP}, \mL^{\dmKP}\}
,\label{equ2}\\
\frac{\pa \log \mP^{\dmKP}}{\pa t_n}=\frac{\pa \B_n^{\dmKP} }{\pa
s} - \{\log\mP^{\dmKP}, \B_n^{\dmKP}\},\label{equ3}
\end{gather}for the power series
\begin{gather}
 \mL^{\dmKP}(k;s,t) = k +u^{\dmKP}_2(s,t)k^{-1}
+u^{\dmKP}_3k^{-2}+\cdots\label{definemL1},\\
\B_n^{\dmKP}(k;s,t)=(\mL^{\dmKP}(k;s,t)^n)_{\geq 0}\nonumber,\\
\mP^{\dmKP}(k;s,t) =k^N + q_1(s,t)k^{N-1}
+\cdots+q_{N}(s,t).\label{definemP1}
\end{gather}
We have
\begin{proposition}
If $(\mL(s), \mP(s))$ is a solution of the dcmKP hierarchy, then the
pair $(\mL^{\dcmKP}(s)$, $ \mP^{\dcmKP}(s))$, where
\begin{gather}\label{def}
\mL^{\dmKP}(s)=\mL^{\dmKP}(k;s,t) = \mL(k-u_1(s,t);s,t), \\
\mP^{\dmKP}(s)=\mP^{\dmKP}(k;s,t)=p_0(s,t)^{-1}\mP(k-u_1(s,t);s,t),\nonumber
\end{gather}
is a solution of the dmKP hierarchy.
\end{proposition}
\begin{proof}
It is easy to see that $\mL^{\dmKP}$ and $\mP^{\dmKP}$ def\/ined by
\eqref{def} has the form required by \eqref{definemL1} and
\eqref{definemP1}. Let $\exp \left(\ad
\sum\limits_{n=0}^{\infty}\phi_n(s,t) k^{-n}\right)$ be the dressing
operator of the solution $(\mL(s), \mP(s))$. Using Corollary
\ref{phi}, it is easy to check that $\mL^{\dmKP}$, $\log\mP^{\dmKP}$
 can be written as
\begin{gather}\label{al}
\mL^{\dmKP} = e^{-\ad \phi_0} \mL, \qquad \log\mP^{\dmKP}=
e^{-\ad\phi_0}\log\mP-\frac{\pa\phi_0}{\pa s}.
\end{gather} Since
\begin{gather*}
e^{-\ad\phi_0}(\mL^n)_{> 0}+ (\mL^n)_{
0}=e^{-\ad\phi_0}(\mL^n)_{\geq 0} = (e^{-\ad\phi_0}\mL^n)_{\geq 0},
\end{gather*} we have
\begin{gather}\label{rel1}
\B_n^{\dmKP} = e^{-\ad\phi_0}\B_n -\frac{\pa \phi_0}{\pa t_n}.
\end{gather} Using this relation, equation \eqref{al}, 
equations \eqref{equation1}--\eqref{equation3} and Lemma A.1 in \cite{TT1}, it is a direct
computation to verify that $(\mL^{\dmKP}, \mP^{\dmKP})$ satisfy
equations \eqref{equ1}--\eqref{equ3}.
\end{proof}

The map \eqref{def} is called a dispersionless Miura map,
corresponding to the Miura map between a solution of KdV hierarchy and a
solution of modif\/ied KdV hierarchy.

\subsection[The special case $\mP=p_0k$]{The special case $\boldsymbol{\mP=p_0k}$}
In the special case where $N=1$ and $\mP=p_0k$, we have from
Corollary \ref{phi} and equation \eqref{Gr2}, \eqref{re2},
\begin{gather*}
\log\mP=\log p_0 + \log k = \frac{\pa\phi_0}{\pa s} +\log \mL
+\sum_{n=1}^{\infty}\frac{\pa\phi_0}{\pa t_n}\mL^{-n}.
\end{gather*}
Comparing with \eqref{P}, we f\/ind that
\begin{gather*}
\frac{\pa^2\log\tau_{\dcmKP}}{\pa t_n\pa s}=\frac{\pa v_n}{\pa s} =
-\frac{\pa \phi_0}{\pa t_n}.
\end{gather*}
Therefore we can f\/ix the dependence of $\tau_{\dcmKP}$ on $s$ by the
equation
\[
    \frac{\pa\log \tau_{\dcmKP}}{\pa s} =-\phi_0
\]
and equation \eqref{dHirota} can be written as
\begin{gather}
z_1 \exp\left(-\sum_{n=1}^{\infty}\frac{1}{n}\frac{\pa^2\F}{\pa
s\pa t_n}z_1^{-n}\right)-z_2
\exp\left(-\sum_{n=1}^{\infty}\frac{1}{n}\frac{\pa^2\F}{\pa s\pa
t_n}z_2^{-n}\right)\nonumber\\
\qquad{}=(z_1-z_2)
  \exp\left(
   \sum_{m=1}^{\infty}\sum_{n=1}^{\infty}
   \frac{1}{mn}\frac{\pa^2\F}{\pa t_m\pa t_n}z_1^{-m}z_2^{-n}
  \right),\label{Hi3}
\end{gather}
which we call the dispersionless Hirota equation for dcmKP hierarchy
with $\mP=p_0k$. The counterpart of Proposition \ref{prop2} becomes
\begin{proposition}
If $\tau_{\dcmKP}(s,t)$ is a function that satisfies the dispersionless Hirota equation
 \eqref{Hi3}, then the pair of
functions $(\mL ,\mP )$, where $\mL(k)=\mL(k;s,t)$ is defined by
inverting
\begin{gather*}
\mathsf{k}(z)=\mathsf{k}(z;s,t)
=z\exp\left(-\sum_{n=1}^{\infty}\frac{1}{n}\frac{\pa^2\log\tau_{\dcmKP}}{\pa
t_n \pa s}z^{-n}\right)
\end{gather*}and  
\[
\mP(k)=\mP(k;s,t)=k \exp\left(-\frac{\pa^2\log\tau_{\dcmKP}}
{\pa s^2}\right)
\]  satisfy the dcmKP hierarchy
\eqref{equation1}--\eqref{equation3}.
\end{proposition}
\begin{proof}
 This can be directly deduced from Proposition~\ref{prop2}.
\end{proof}

Comparing with Proposition 3.1 in \cite{Teo2} and its following
discussion, we f\/ind that if $(\mL(k;s,t)$, $\mP(k;s,t) = p_0(s,t)k)$ is
a solution of dcmKP hierarchy, $\mL(k;s,t)$ is a solution of the
hierarchy
\begin{gather*}
\frac{\pa \mL}{\pa t_n} =\{ (\mL^n)_{\geq 0},\mL\}_T;
\end{gather*} $\mL_1(k;s,t)= \mL(p_0^{-1}k;s,t)$ is a solution of
the hierarchy
\begin{gather}\label{v2}
\frac{\pa \mL_1}{\pa t_n} =\{ (\mL_1^n)_{> 0},\mL_1\}_T,
\end{gather} and $\mL_{1/2}(k;s,t)= \mL(p_0^{-1/2}k;s,t)$ is a
solution of the hierarchy
\begin{gather}\label{v3}
\frac{\pa \mL_{1/2}}{\pa t_n} =\left\{ (\mL_{1/2}^n)_{>
0}+\frac{1}{2}(\mL_{1/2}^n)_0,\;\mL_{1/2}\right\}_T.
\end{gather}
Here $\{\cdot, \cdot\}$ is the Poisson bracket of dToda hierarchy:
\[
\{ f(k,s), g(k,s)\}_T = k\frac{\pa f}{\pa k}\frac{\pa g}{\pa s}
-k\frac{\pa f}{\pa s}\frac{\pa g}{\pa k}.
\]
\eqref{v2} and \eqref{v3}
can be considered as gauge equivalent form of the dcmKP hierarchy
with $\mP=p_0k$ and  with gauge parameter $1$ and $1/2$
respectively. For the cmKP version of \eqref{v2} and~\eqref{v3} we refer
to Appendices \ref{app:diff-ce}, \ref{app:cmkp:gauge} and~\ref{app:diff-ce=diff-l}. 
The form \eqref{v3} was used in the work~\cite{tak-tak:05}.

\subsection{Quasi-classical limit of the cmKP hierarchy}
\label{subsec:cmkp->dcmkp}
The dispersionless KP hierarchy and the dispersionless Toda hierarchy
are obtained as quasi-classical limit of corresponding ``dispersionful''
hierarchies. See \cite{TT1}. The dcmKP hierarchy is also quasi-classical
limit of the cmKP hierarchy. In this subsection we brief\/ly explain the
correspondence.

Let us def\/ine the {\it order} as in \cite{TT1}, \S~1.7.1:
\begin{gather}
    \ord^\hbar \left( \sum a_{n,m}(t) \hbar^n \der^m \right)
    :=
    \max \{ m-n \,|\, a_{n,m}(t) \neq 0 \}.
\label{def:ord} 
\end{gather}
In particular, $\ord^\hbar (\hbar) = -1$, $\ord^\hbar(\der) = 1$.
The {\it principal symbol} (resp.\ the {\it symbol of order $l$})
of an operator $A = \sum a_{n,m}\hbar^n \der^m$ is
\begin{gather*}
    \symbolh (A)
   := \hbar^{-\ord^\hbar(A)} \!\! \sum_{m-n = \ord^\hbar(A)} a_{n,m}  k^m ,
\end{gather*}
respectively
\begin{gather*} 
    \symbolh_l(A)
    := \hbar^{-l} \!\! \sum_{m-n = l} a_{n,m}  k^m.
\end{gather*}

Let us redef\/ine the cmKP hierarchy with a small parameter $\hbar$ as
follows. Fix a positive integer $N$. The discrete independent variable
$s$ runs in $S:=\hbar\Integer$. Operators $L$, $P$ are of the form
\begin{gather}
    L(s;x,t)
    = \sum_{n=0}^\infty u_n(\hbar,s,x,t) (\hbar\der)^{1-n},
\label{def:L(s):h}
\\
    P(s;x,t)
    = \sum_{n=0}^{N-1} p_n(\hbar,s,x,t) (\hbar\der)^{N - n},
\label{def:P(s):h}
\end{gather}
where $u_0=1$, $p_0\neq 0$ and all coef\/f\/icients $u_n(\hbar,s,x,t)$ and
$p_n(\hbar,s,x,t)$ are regular in $\hbar$. Namely they do not contain
negative powers of $\hbar$. The cmKP hierarchy is rewritten as
\begin{gather}
    \hbar\frac{\der L(s)}{\der t_n} = [B_n(s), L(s)],
\label{cmkp:dL/dt:h}
\\
    (L(s+\hbar) - L(s)) P(s)= [P(s), L(s)],
\label{cmkp:LP=PL:h}
\\
    \left(B_n(s) - B_n(s+\hbar) \right) P(s)
    =
    \left[P(s),
          \hbar\frac{\der}{\der t_n} - B_n(s) \right].
\label{cmkp:BP=PB:h}
\end{gather}
where $B_n(s) = B_n(s;x,t)$ is def\/ined as before. The principal symbols
of \eqref{cmkp:dL/dt:h}, \eqref{cmkp:LP=PL:h} and \eqref{cmkp:BP=PB:h}
give equations \eqref{equation1}, \eqref{equation2} and
\eqref{equation3} of the dcmKP hierarchy respectively, where
$\mL(s)=\symbolh_0(L(s))$ and $\mP(s)=\symbolh_0(P(s))$.

The dressing operator $W(s;x,t;\der)$ \eqref{dressing-op:diff-l} should
have the form
\begin{gather*}
    W(s;x,t;\hbar;\der)
    =
    \exp(\hbar^{-1}X(\hbar,x,t;\der)) (\hbar\der)^{Ns},
\\
    X(\hbar,x,t;\der)
    = \sum_{n=0}^\infty \chi_n(\hbar,x,t) (\hbar\der)^{-n},
\end{gather*}
where $\chi_n(\hbar,x,t)$ is regular in $\hbar$. The principal
symbol of $X$ is the function $\phi$ in \propref{varphi}.

\begin{remark}
\label{lift:dcmkp->cmkp}
 Solutions of the dispersionless KP and Toda hierarchies can be lifted
 up to solutions of the KP and Toda hierarchies (with $\hbar$)
 respectively by lifting the dressing operator. See Corollary 1.7.6 and
 Corollary 2.7.6 of \cite{TT1}. 

 As for the dcmKP hierarchy, we conjecture that any solution of the
 dcmKP hierarchy would be lifted up to a solution of the cmKP hierarchy
 but there is dif\/f\/iculty coming from the form of~$P(s)$. Naive lift of
 $\phi(s,t)$ would cause a tail of~$P(s)$ which has non-positive order
 as a micro-dif\/ferential operator and negative order in the sense of
 \eqref{def:ord}. We can correct this by inductively modifying $X(s)$ to
 get $P(s)$ of the form \eqref{def:P(s):h}, but this inductive procedure
 might make $X(s)$ behave wildly with respect to $s$.
\end{remark}

In the context of the WKB analysis of the linear equations
\eqref{lin-eq:wave-func} the $S$ function introduced in
\secref{subsec:2-form-and-S} is the phase function:
\begin{gather}
    w(s;x,t;\lambda)
    = \exp\bigl(\hbar^{-1} (S + O(\hbar))\bigr),
\label{w=exp(S/h)}
\\
    S= \sum_{n=1}^{\infty} t_n \lambda^n +x\lambda + N s\log \lambda
     - \sum_{n=1}^{\infty} \frac{v_n}{n}\lambda^{-n} +\phi_0.
\label{S:wkb}
\end{gather}
By replacing $\lambda$ with $\mL$, we obtain the $S$ function in
\propref{prop:S}.

If the conjecture in \remref{lift:dcmkp->cmkp} is true, the form of
$\mP(s)$ of the dcmKP hierarchy reduces drastically by the following
proposition.
\begin{proposition}
\label{prop:P(s)=p0kN:dcmkp}
 If a solution $(\mL(s),\mP(s))_{s\in S}$ of the dcmKP hierarchy is the
 quasi-classical limit of a solution $(L(s),P(s))_{s\in S}$ of the cmKP
 hierarchy with $\hbar$. Then $\mP(s) = p_0(s,t) k^N$. 
\end{proposition}

\begin{proof}
 Let $\tau_0(s;t)$ and $\tau_1(s;t)$ be the tau function of
 $(L(s),P(s))_{s\in S}$. They are expressed by the Clif\/ford algebra as
 in \eqref{def:tau0,tau1}: for $s\in\Integer\hbar$
\begin{gather}
    \tau_0(\hbar;s;t) 
    =
    \langle Ns\hbar^{-1} | e^{J(t)\hbar^{-1}} g 
          | Ns\hbar^{-1} \rangle, \nonumber
\\
    \tau_1(\hbar;s;t) 
    =
    \langle Ns\hbar^{-1} + 1| e^{J(t)\hbar^{-1} } g 
          | Ns\hbar^{-1} + 1 \rangle.
\label{def:tau0,tau1:h}
\end{gather}

 Def\/ining
\begin{gather}
    \tau(s;t)
    := \langle s\hbar^{-1}|e^{J(t)\hbar^{-1}}g|s\hbar^{-1}\rangle
\label{tau-interpolated}
\end{gather}
 we have a solution $(\tilde L(s), \tilde P(s))_{s\in\Integer}$ of the
 cmKP hierarchy with $N=1$ whose $\tau$ functions are
 $\tilde\tau_0(s;t)=\tau(s;t)$, $\tilde\tau_1(s;t)=\tau(s+\hbar;t)$. It
 is easy to see that the dressing operators $W(s)$ of $(L(s),P(s))_{s\in
 S}$ and $\tilde W(s)$ of $(\tilde L(s),\tilde P(s))_{s\in\Integer}$ are
 related by $W(s)=\tilde W(Ns)$.

 Hence $L(s) = \tilde L(Ns)$ and $P(s)=\tilde
 P(Ns+(N-1)\hbar)\cdots\tilde P(Ns)$. The symbol of order $0$ of the
 last equation gives
\begin{gather*}
    \mP(s) = \symbolh_0(P(s))
    = \symbolh_0(\tilde P(Ns+(N-1)\hbar)) \cdots 
      \symbolh_0(\tilde P(Ns)),
\end{gather*}
 because $\symbolh_0(AB)=\symbolh_0(A)\symbolh_0(B)$ for any operators
 $A,B$, $\ord^\hbar(A)=\ord^\hbar(B)=0$. Since $\tilde P(s)$ is of the form
 $\tilde p_0(s,t)\hbar\der$, $\mP(s)$ is of the form $p_0(s,t) k^N$.
\end{proof}

A proof of the above statement without lifting up to the cmKP hierarchy
is desirable.

{\samepage If $\mP(s) = p_0(s,t)k^N$, we have $\Res \mL^n d_k \log \mP(s) = N
(\mL^n)_0$, which is equivalent to
\begin{gather}
    \frac{\der v_n}{\der s} = -N \frac{\der \phi_0}{\der t_n},
\label{dvn/ds=-Ndphi0/dtn}
\end{gather}
because of \eqref{P} and Corollary \ref{phi}. This equation together
with \eqref{sym1} is a compatibility condition of equations \eqref{tau}
and 
\begin{gather}
    \frac{\der\log\tau_{\dcmKP}}{\der s} = -N\phi_0,
\label{dtau/ds}
\end{gather}
which f\/ixes the $s$-dependence of $\log\tau_{\dcmKP}$. }

In fact, this is consistent with the quasi-classical limit. We express
the $\tau$ functions of $(L(s),P(s))_{s\in S}$ as
\eqref{def:tau0,tau1:h} and def\/ine $\tau(s;t)$ by
\eqref{tau-interpolated}. As in \cite{TT1}, that $\tau$ function behaves
as
\begin{gather*}
    \tau(s;t) = e^{\hbar^{-2} F(\hbar,s,t)}
\end{gather*}
and therefore $\tau_0$ and $\tau_1$ behave as
\[
    \tau_0(\hbar;s;t) = e^{\hbar^{-2}F(\hbar,Ns,t)}, \qquad
    \tau_1(\hbar;s;t) = e^{\hbar^{-2}F(\hbar,Ns+\hbar,t)}.
\]
Substituting this into \eqref{w=tau/tau} and comparing the result with
\eqref{w=exp(S/h)} and \eqref{S:wkb}, we have
\[
    \frac{\der F(\hbar,Ns,t)}{\der t_n} = v_n(s,t) + O(\hbar),
\qquad
    - \frac{\hbar}{N} \frac{\der F(\hbar,Ns,t)}{\der s}
    = \hbar \phi_0(s,t) + O(\hbar^2).
\]
Hence $\log\tau_{\dcmKP}(s;t) = F(\hbar,Ns,t)|_{\hbar=0}$, which
satisf\/ies \eqref{tau} and \eqref{dtau/ds}.

\appendix
\section[Form of $P(s)$, $\mP(s)$ and proof of
\propref{prop:cmkp=mkp+gauge} (\hbox{\rm ii})]{Form of $\boldsymbol{P(s)}$, $\boldsymbol{\mP(s)}$ and proof of
\propref{prop:cmkp=mkp+gauge} (ii)}
\label{app:P(s)}
In the main text we assumed that operator $P(s)$ of the cmKP hierarchy
does not have the $0$-th order term as in \eqref{def:P(s)}. We also put
similar requirement \eqref{definemP} to $\mP(s)$ of the dcmKP
hierarchy. At f\/irst glance, these assumptions might seem artif\/icial but
in fact they are not restriction as we show in this appendix.

Assume that $L(s)$ is of the form \eqref{def:L(s)} and that $P(s)$
has the form
\begin{gather}
    P(s;x,t)
    := p_0(s,x,t) \der^{m_s}
     + \cdots + p_{m_s-1}(s,t) \der + p_{m_s}(s,x,t)
\nonumber\\
\phantom{P(s;x,t)}{}= \sum_{n=0}^{m_s} p_n(s,x,t) \der^{m_s - n}, \qquad
    p_0(s,x,t) \neq 0,
\label{def:P:0}
\end{gather}
instead of the form \eqref{def:P(s)}. Assume further that
$(L(s),P(s))_{s\in S}$ satisf\/ies the system \eqref{cmkp:dL/dt},
\eqref{cmkp:LP=PL} and \eqref{cmkp:BP=PB}. We show that there exists
a function $f(s)=f(s,x,t)$ which satisf\/ies
\begin{itemize}\itemsep=0pt 
 \item The pair $(\tilde L(s) := f(s)^{-1} L(s) f(s), \tilde P(s) :=
       f(s+1)^{-1} P(s) f(s))_{s\in S}$ is a solution of the cmKP
       hierarchy.
 \item $\tilde P(s)$ does not have the $0$-th order term:
\begin{gather}
    \tilde P(s) = \tilde p_0(s) \der^{m_s}
    + \cdots + \tilde p_{m_s-1}(s) \der + \tilde p_{m_s}(s), \qquad
    \tilde p_{m_s}(s)=0.
\label{tildeP}
\end{gather}
\end{itemize}
In this sense we can assume without loss of generality that
$p_{m_s}(s)=0$ in \eqref{def:P(s)}.

The following is the basic lemma:
\begin{lemma}
\label{lem:0-th-term}
 Let $Q = \sum\limits_{j=0}^N q_j(x) \der^{N-j}$ be a differential operator and
 $f(x)$ is a function. Then the $0$-th order term of the composition
 $Q\circ f$ is the function $Q(f)$ obtained by applying $Q$ on $f$.
\end{lemma}
This is a direct consequence of the Leibniz rule: $\der^k \circ f =
\sum\limits_{r=0}^k \binom{k}{r} f^{(k-r)} \der^r$.

Hence the second condition \eqref{tildeP} for the function $f(s,t)$
is equivalent to
\begin{gather}
    P(s)(f(s)) = 0.
\label{Pf=0}
\end{gather}

Let us introduce an operator $C_n(s)$ by
\begin{gather}
    f(s)^{-1} (\der_{t_n} - B_n(s)) f(s)
    =
    \der_{t_n} - C_n(s),
\nonumber\\ 
\text{i.e.,}\quad 
    C_n(s) = f(s)^{-1} B_n(s) f(s) - f(s)^{-1} \frac{\der f(s)}{\der t_n}.
\label{Cn}
\end{gather}
Then, it follows from \eqref{cmkp:dL/dt}, \eqref{cmkp:LP=PL} and
\eqref{cmkp:BP=PB} that the pair $(\tilde L(s), \tilde P(s))_{s\in
S}$ satisf\/ies the following equations:
\begin{gather}
    [\tilde L(s), \der_{t_n} - C_n(s)] = 0,
\label{cmkp:lax:tildeL}
\\
    \tilde L(s+1) \tilde P(s) = \tilde P(s) \tilde L(s),
\label{cmkp:tildeLP=PL}
\\
    (\der_{t_n} - C_n(s+1)) \tilde P(s)
    = \tilde P(s) (\der_{t_n} - C_n(s)).
\label{cmkp:tildeCP=PC}
\end{gather}
Hence if $C_n(s) = (\tilde L(s)^n)_{>0}$, then $(\tilde L(s), \tilde
P(s))$ is a solution of the cmKP hierarchy. By
\lemref{lem:0-th-term} we have
\begin{gather*}
    C_n(s) = (f(s)^{-1} B_n(s) f(s))_{>0}
           + f(s)^{-1} B_n(s)(f(s))
           - f(s)^{-1} \frac{\der f(s)}{\der t_n}.
\end{gather*}
Therefore the condition $C_n(s) = (\tilde L(s)^n)_{>0}$ is
equivalent to
\begin{gather}
    (\der_{t_n} - B_n(s)) (f(s)) = 0.
\label{(d-Bn)f=0}
\end{gather}

So, we have to f\/ind a function $f(s)$ satisfying \eqref{Pf=0} and
\eqref{(d-Bn)f=0}. This is done inductively as follows. First solve
equation \eqref{Pf=0} for $t_1=t_2=\cdots=0$ but for arbitrary $x$.
We denote the solution by $f_0(s,x)$:
\begin{gather}
    P(s;x, t_1=t_2=\cdots=0) f_0(s,x) = 0, \qquad
    f_0(s,0)=1.
\label{f0}
\end{gather}
Function $f_1(s,x,t_1) = f_0(s,x+t_1)$ satisf\/ies \eqref{Pf=0} as
well as \eqref{(d-Bn)f=0} with $n=1$ for $t_2=t_3=\cdots=0$.

Suppose we have function $f_m(s,x,t_1,\dots,t_m)$ which satisf\/ies
\eqref{Pf=0} and \eqref{(d-Bn)f=0} with $n=1,\dots,m$ and $t_k=0$
($k>m$). We can solve the Cauchy problem
\begin{gather}
\frac{\der}{\der t_{m+1}} f_{m+1}(s,x,t_1,\dots,t_{m+1})
\nonumber\\
    \qquad{}- B_{m+1}(s,x,t_1,\dots,t_{m+1}, 0,0,\dots)
    f_{m+1}(s,x,t_1,\dots,t_{m+1}) = 0,\label{fm+1}\\
f_{m+1}(s,x,t_1,\dots,t_m,0) = f_m(s,x,t_1,\dots,t_m),\nonumber
\end{gather}
with respect to $t_{m+1}$. By \eqref{cmkp:zs} and \eqref{cmkp:BP=PB}
for $n=m+1$, the solution $f_{m+1}$ of \eqref{fm+1} satisfies~\eqref{Pf=0} 
and \eqref{(d-Bn)f=0} for $n=1,\dots,m+1$ and $t_k=0$
($k>m+1$).

The desired function $f(s,t)=f(s,x,t_1,t_2,\dots)$ is def\/ined by the
inductive limit of the sequence $f_n(s,x,t_1,t_2,\dots,t_n)$.

The second statement of \propref{prop:cmkp=mkp+gauge} is proved in
the same way. Suppose that a solution $(L^\mkp(s),P^\mkp(s))_{s\in
S}$ of the mKP hierarchy and a sequence
$\{(f^{(0)}(s),\dots,f^{(m_s-1)}(s)) \}_{s\in S}$ of non-zero
constant vectors are given. Replace $L(s)$, $P(s)$ and $B_n(s)$ in
the above argument by $L^\mkp(s)$, $P^\mkp(s)$ and $B^\mkp_n(s)$
respectively. (See \eqref{def:Lmkp}, \eqref{def:Pmkp} and
\eqref{def:Bmkp}.) If we solve equation \eqref{f0} under the initial
condition
\begin{gather*}
    \der^k f_0(s,0) = f^{(k)}(s), \qquad k=0,\dots,m_s-1, \quad (s\in S'),
\end{gather*}
we obtain a function $f(s)=f(s,x,t)$ such that
\begin{gather*}
    \der^k f_0(s;0,0) = f^{(k)}(s)\qquad
    \text{for all}\quad s \in S, \quad 0 \leqq k < m_s,
\end{gather*}
and $(L(s):= f(s)^{-1} L^\mkp(s) f(s), P(s):= f(s+1)^{-1}
P^\mkp(s)f(s))_{s\in S}$ is a solution of the cmKP hierarchy.
\qed

We proceed to the case of the dcmKP hierarchy.

\begin{lemma}
\label{lem:0-th-term:dcmkp}
  Let $(\mL(s), \mP(s))$, where
\begin{gather}
  \mL(s)=\mL(k;s,t) = k +\sum_{n=0}^{\infty} u_{n+1}(s,t)k^{-n},\nonumber\\
  \mP(s)=\mP(k;s,t) = p_0(s,t)k^N+\cdots + p_N(s,t),
 \label{form}
\end{gather}
be a solution of the dcmKP hierarchy.  If $\varphi(s,t)$ is a function
that satisfies the system of equations\footnote{The equation when $n=1$
is a tautology.}
\begin{gather}
  \frac{\pa\varphi(s,t)}{\pa t_n}
  =
  -\B_n\left(-\frac{\pa\varphi(s,t)}{\pa x}; s,t\right),
  \qquad n\geq1,
\label{gauge-away-func:dcmkp}
\end{gather}
then the pair $(\tilde{\mL}(s), \tilde{\mP}(s))$, where
\begin{gather*}
\tilde{\mL}(k;s,t) = e^{\ad\varphi(s,t)}\mL(k;s,t)
=\mL\left(k-\frac{\pa \varphi(s,t)}{\pa x} ; s,t\right),\\
\tilde{\mP}(k;s,t) = e^{\pa\varphi(s,t)/\pa
s}e^{\ad\varphi(s,t)}\mL(k;s,t) =e^{\pa\varphi(s,t)/\pa
s}\mP\left(k-\frac{\pa \varphi(s,t)}{\pa x} ; s,t\right),
\end{gather*}is also a solution of the dcmKP hierarchy.
\end{lemma}

\begin{proof}
First, observe that
\[
 \tilde{\B}_n = (\tilde{\mL}^n)_{>0} =
 e^{\ad\varphi}\B_n-\B_n\left(-\frac{\pa\varphi}{\pa x}\right).
\]
Therefore, we have
\begin{gather*} 
  \frac{\pa\tilde{\mL}}{\pa t_n}
  =\left\{\frac{\pa\varphi}{\pa t_n}, \tilde{\mL}\right\} +
  e^{\ad\varphi}\{\B_n, \mL\}\\
  \phantom{\frac{\pa\tilde{\mL}}{\pa t_n}}{}=
  \left\{-\B_n\left(-\frac{\pa\varphi}{\pa
  x}\right)+e^{\ad\varphi}\B_n, \tilde{\mL}\right\}
  =\{\tilde{\B}_n,\tilde{\mL}\}.
\end{gather*}
Similarly,
\begin{gather*}
\frac{\pa\tilde{\mL}}{\pa s} =\left\{\frac{\pa\varphi}{\pa s}
+e^{\ad\varphi}\log\mP,
\tilde{\mL}\right\}=\{\log\tilde{\mP},\tilde{\mL}\}.
\end{gather*}Finally,
\begin{gather*}
\frac{\pa \log \tilde{\mP}}{\pa t_n} =\frac{\pa^2\varphi}{\pa
t_n\pa s} +\left\{ \frac{\pa\varphi}{\pa t_n},
e^{\ad\varphi}\log\mP\right\} +e^{\ad
\varphi}\left(\frac{\pa\B_n}{\pa s} -\{\log \mP, \B_n\}\right)\\
\phantom{\frac{\pa \log \tilde{\mP}}{\pa t_n}}{}=\left\{ \frac{\pa\varphi}{\pa
t_n},\log\tilde{\mP}\right\}+\frac{\pa^2\varphi}{\pa s\pa
t_n}+\frac{\pa }{\pa s}\left(e^{\ad\varphi}\B_n\right)
-\left\{\frac{\pa\varphi}{\pa s}, e^{\ad\varphi}\B_n\right\}\\
\phantom{\frac{\pa \log \tilde{\mP}}{\pa t_n}=}{}-\left\{e^{\ad\varphi}\log\mP, e^{\ad\varphi}\B_n\right\}\\
\phantom{\frac{\pa \log \tilde{\mP}}{\pa t_n}}{}=\frac{\pa \tilde{\B}_n}{\pa s}-\left\{\log\tilde{\mP},
\tilde{\B}_n\right\}.\tag*{\qed}
\end{gather*}\renewcommand{\qed}{}
\end{proof}

\begin{lemma}
\label{lem:0-th-term2:dcmkp}
 Let $(\mL(s), \mP(s))$ be as in the lemma above. If $\mP(k;s,t)$ has a
 root $\psi(s,t)$ as a~polynomial of $k$, the system
\begin{gather}
\label{eq1}
 \frac{\pa\varphi(s,t)}{\pa t_n} =-\B_n (\psi(s,t); s,t),
\qquad n\geq 1,
\end{gather}
 has a solution, unique up to a function of $s$.
\end{lemma}

In particular, \eqref{eq1} for $n=1$ implies that $\psi(s,t) =
-\frac{\der \varphi}{\der x}$. Hence the function satisfying
\eqref{gauge-away-func:dcmkp} is obtained and we can gauge away the
constant term of $\mP(s)$ according to \lemref{lem:0-th-term:dcmkp}. 

\begin{proof}
 Let us factorize $\mP(k;s,t)$ as $\mP(k;s,t) = p_0(s,t)
 \prod\limits_{i=1}^N(k-\psi_i(s,t))$. Dif\/ferentiating by $t_n$, we have
\begin{gather}
    \frac{\der \log \mP}{\der t_n}
    = 
    \sum_{i=1}^N
    \frac{-\frac{\der \psi_i}{\der t_n}}{k-\psi_i}
    + \frac{\der}{\der t_n} \log p_0.
\label{dlogP/dtn}
\end{gather}
 The left hand side is, due to \eqref{equation3},
\begin{gather*}
    \frac{\der \log \mP}{\der t_n}
    = 
    \frac{\der \B_n}{\der s}
    - \frac{\der \log \mP}{\der k} \frac{\der \B_n}{\der x}
    + \frac{\der \log \mP}{\der x} \frac{\der \B_n}{\der k}
\\
    \phantom{\frac{\der \log \mP}{\der t_n}}{}= 
    \frac{\der \B_n}{\der s}
    - \frac{\der \B_n}{\der x} \sum_{i=1}^N \frac{1}{k-\psi_i}
    + \frac{\der \B_n}{\der k} 
      \sum_{i=1}^N \frac{-\frac{\der\psi_i}{\der x}}{k-\psi_i}.
\end{gather*}
 Substituting this into \eqref{dlogP/dtn}, multiplying
 $\frac{\der\B_m}{\der k}$, subtracting the same equation with $m$ and
 $n$ interchanged, we have
\begin{gather*}
    \{\B_n, \B_m\}(\psi_i;s,t)
    =
    \frac{\der\B_n}{\der k}(\psi_i;s,t) 
    \frac{\der\psi_i(s,t)}{\der t_m}
    -
    \frac{\der\B_m}{\der k}(\psi_i;s,t)
    \frac{\der\psi_i(s,t)}{\der t_n}
\end{gather*}
 by comparing the residue at $k=\psi_i$. Using this equation, we can
 check the consistency of the system \eqref{eq1} as follows:
\begin{gather}
    \frac{\pa}{\pa t_m} \B_n(\psi(s,t);s,t) 
    - \frac{\pa}{\pa t_n}\B_m(\psi(s,t);s,t)
\nonumber\\
 \qquad{}   =
    \left(\frac{\pa\B_n}{\pa t_m} -\frac{\pa\B_m}{\pa t_n}\right)
    (\psi(s,t);s,t)
\nonumber\\
 \qquad\quad{}{}   +\frac{\pa\B_n}{\pa k}(\psi(s,t);s,t)
     \frac{\pa\psi(s,t)}{\pa t_m}
    -\frac{\pa\B_m}{\pa k}(\psi(s,t);s,t)
     \frac{\pa\psi(s,t)}{\pa t_n}
\nonumber\\
 \qquad{}   =
    \left(
     \frac{\pa\B_n}{\pa t_m}-\frac{\pa\B_m}{\pa t_n}
    +\left\{\B_n,\B_m\right\}\right)(\psi(s,t);s,t)=0.
\end{gather}
 Therefore, the system \eqref{eq1} has a solution $\varphi(s,t)$ unique
 up to a function of $s$. 
\end{proof}

\section[Difference operator formalism]{Dif\/ference operator formalism}
\label{app:diff-ce}

When the set $\{n_s\}_{s\in S}$ is equal to the whole set of integer numbers 
$\Integer$, we can formulate the cmKP hierarchy in terms of
dif\/ference operators.  In fact in this case the cmKP hierarchy can
be thought of as the ``half'' of the Toda lattice hierarchy of Ueno
and Takasaki \cite{uen-tak:84} whose dependence on half of the time
variables are suppressed.

In this appendix, we f\/irst present the dif\/ference operator formalism of
the cmKP hierarchy and then show in \appref{app:diff-ce=diff-l} that it
is equivalent to the cmKP hierarchy in the main text. We also introduce
a gauge parameter $\alpha$, $\alpha\neq 0$. (See \cite{tak:90} for the
gauge parameter of the Toda lattice hierarchy.)

Let $\boldL$ be a dif\/ference operator of the form
\begin{gather}
    \boldL
    = b_0(s,t) e^{\der_s} + b_1(s,t) + b_2(s,t) e^{-\der_s} + \cdots
    = \sum_{j=0}^\infty b_j(s,t) e^{(1-j)\der_s},
\label{bL}
\end{gather}
where $e^{k \der_s}$ is the $k$-step shift operator $e^{k\der_s}f(s)
= f(k+s)$ and $t=(t_1,t_2,t_3,\dots)$ is a sequence of continuous
variables. We assume that $b_0$ never vanishes: $b_0(s,t) \neq 0$.

We call the following system the {\em difference operator formalism
of the cmKP hierarchy} with gauge parameter $\alpha$:
\begin{gather}
    \frac{\der \boldL}{\der t_n} = [\boldB_n, \boldL].
\label{cmkp:diff-ce}
\end{gather}
Here $\boldB_n$ is a dif\/ference operator def\/ined by
\begin{gather}
    \boldB_n := (\boldL^n)_{\geq 0} - \alpha (\boldL^n)_0,
\label{bB}
\end{gather}
where $(\cdot)_{\geq 0}$ and $(\cdot)_0$ are projections of
dif\/ference operators: for $A = \sum\limits_{j} a_j(s) e^{j\der_s}$,
\begin{gather}
    A_{\geq 0} := \sum_{j\geq 0} a_j(s) e^{j\der_s}, \qquad
    A_{0} := a_0(s), \qquad
    A_{< 0} := \sum_{j<0} a_j(s) e^{j\der_s}.
\label{def:proj:diff-ce}
\end{gather}

By the same argument as in \cite[\S~1]{uen-tak:84}, the Lax
representation \eqref{cmkp:diff-ce} of the cmKP hierarchy is
equivalent to the Zakharov--Shabat (or zero-curvature)
representations:
\begin{gather}
    [\der_{t_m} - \boldB_m, \der_{t_n} - \boldB_n] = 0,
\label{cmkp:diff-ce:zs}
\\
    [\der_{t_m} - \boldB^c_m, \der_{t_n} - \boldB^c_n] = 0,
\label{cmkp:diff-ce:zs-c}
\end{gather}
where
\begin{gather}
     \boldB^c_n = \boldB_n - \boldL^n
     = - (\boldL^n)_{<0} - \alpha (\boldL^n)_0.
\label{bBc}
\end{gather}

\begin{example}
\label{ex:a=1/2:diff-ce}
 Dispersionless limit of the case $\alpha=1/2$ is \eqref{v3}. It is
 related to the L\"owner equation. See \cite{tak-tak:05}.
\end{example}

The proof of the following proposition is the same as those of
Theorem 1.2 of \cite{uen-tak:84}.
\begin{proposition}
\label{prop:linear-problem}
 For each solution of the cmKP hierarchy with gauge parameter $\alpha$,
 there exists a difference operator $\hat\boldW$ of the following form
 with coefficients $w_0(s,t)=e^{-\alpha \varphi(s,t)}$ and $w_j(s,t)$:
\begin{gather}
    \hat\boldW
    = e^{-\alpha \varphi(s,t)} + w_1(s,t) e^{-\der_s} + \cdots
    = \sum_{j=0}^\infty w_j(s,t) e^{-j\der_s},
\label{dressing-operator}
\end{gather}
 satisfying the equations
\begin{gather}
   \boldL = \hat\boldW e^{\der_s} \hat\boldW^{-1}, \qquad
   \frac{\der \hat\boldW}{\der t_n} = \boldB^c_n \hat\boldW,
\label{lin-eq:diff-ce}
\end{gather}
 where the operators $\boldB^c_n$ are defined by \eqref{bBc}.
\end{proposition}

We call $\hat\boldW$ the {\em wave matrix}.

\begin{proposition}
\label{prop:bilinear:wave:diff-ce}
 (i)
 The operator $\hat\boldW$ in \propref{prop:linear-problem} and the
 function $\varphi(s,t)$ satisfy the following bilinear equation for
 arbitrary $t$ and $t'$:
\begin{gather}
    \boldW(t) \boldW(t')^{-1}
    =
    e^{(1-\alpha) (\varphi(s,t) - \varphi(s,t'))}
    + (\text{\rm strictly upper triangular}),
\label{bilinear:bW}
\end{gather}
 where $\boldW(t) = \hat\boldW(t) \exp\left(\sum\limits_{n=1}^\infty t_n
 e^{n\der_s}\right)$ and the ``{\rm (strictly upper triangular)}'' part
 is an operator of the form $\sum\limits_{n>0} a_n(s) e^{n\der_s}$.

 (ii)
 Conversely, if a function $\varphi(s,t)$ and an operator $\hat\boldW$
 of the form \eqref{dressing-operator} satisfies the equation
 \eqref{bilinear:bW}, then the operator $\boldL$ def\/ined by
 $\boldL=\hat\boldW e^{\der_s} \hat\boldW^{-1}$ is a solution of the
 cmKP hierarchy \eqref{cmkp:diff-ce}.
\end{proposition}

The proof is essentially the same as the proof of Proposition 1.4
and Theorem 1.5 of \cite{uen-tak:84}.

By means of $\varphi(s,t)$ in \propref{prop:linear-problem}, we can
change the gauge as follows.

\begin{proposition}
\label{prop:gauge-trans:diff-ce}
 Let $\boldL$ be a solution of the cmKP hierarchy \eqref{cmkp:diff-ce}
 with gauge parameter $\alpha$ and~$\varphi$ be the function defined in
 \propref{prop:linear-problem}. Then the difference operator defined by
\begin{gather}
    \tilde\boldL
    := e^{(\alpha-\beta)\varphi(s,t)} \boldL
       e^{-(\alpha-\beta)\varphi(s,t)}
\label{def:tildebL}
\end{gather}
 satisfies equation \eqref{cmkp:diff-ce} with gauge parameter
 $\beta$. Here $\beta$ can be $0$.
\end{proposition}

Simple calculation is sufficient to check  \eqref{cmkp:diff-ce} for
$\tilde\boldL$. We have only to note that the truncating operations
$(\cdot)_{\geq 0}$, $(\cdot)_{0}$ (cf.\ \eqref{def:proj:diff-ce})
commute with the adjoint operation $e^{(\alpha-\beta)\varphi(s,t)}
(\cdot) e^{-(\alpha-\beta)\varphi(s,t)}$.

\section{The cmKP hierarchy with a gauge parameter}
\label{app:cmkp:gauge}

As is naturally expected from \secref{app:diff-ce}, we can introduce
a gauge parameter $\alpha$ ($\alpha\neq 0$) in the cmKP hierarchy
when $\{n_s\}_{s\in S}=\Integer$. In this case operator $L(s)$ has
the form as in \eqref{def:L(s)} but $P(s)$ has a $0$-th order term:
\begin{gather}
    L(s)=L(t;s)
    := \der + u_1(s,t) + u_2(s,t) \der^{-1} + u_3(s,t) \der^{-2}
     + \cdots,
\label{def:L(s):gauge}
\\
    P(s)=P(s,t)
    := p_0(s,t) \der + p_1(s,t),
\label{def:P:gauge}
\end{gather}
which satisfy the condition
\begin{gather}
    (1-\alpha) p_0(s,t) u_1(s,t) + \alpha p_1(s,t) = 0.
\label{p0,u1,p1}
\end{gather}
This condition is equivalent to saying that $B_1(s)$ def\/ined later
is equal to $\der$. The cmKP hierarchy in \secref{subsec:def-cmkp}
is recovered when $\alpha=1$.

We introduce operators $P^{(n)}(s)$ which play the role of the
$n$-step shift operators:
\begin{gather}
    P^{(n)}(s):= \begin{cases}
    P(s+n-1) \cdots P(s+1) P(s), &n>0,\\
    1, &n=0,\\
    P(s+n)^{-1} \cdots P(s-2)^{-1} P(s-1)^{-1}, &n<0.
    \end{cases}
\label{def:P(n)(s)}
\end{gather}
The fundamental properties of $P^{(n)}(s)$ are the following:

\begin{lemma}
\label{lem:P(n)(s)}
{\rm (i)}
 Any microdifferential operator $Q$ has a unique expansion of the form
\begin{gather}
    Q = \sum_{\nu \in \Integer} a_\nu P^{(\nu)}(s).
\label{expand:P(n)(s)}
\end{gather}
 If $Q$ is a $n$-th order differential operator, the sum is taken over
 $0\leqq \nu \leqq n$.

{\rm (ii)}
\begin{gather}
    P^{(m)}(s+n)\, P^{(n)}(s) = P^{(m+n)}(s).
\label{P(n)(s):composition}
\end{gather}
\end{lemma}

According to \lemref{lem:P(n)(s)} (i), the operator $L(s)^n$ is
expanded as:
\begin{gather}
    L(s)^n = \sum_{j=0}^\infty b^{(n)}_j(s,t) P^{(n-j)}(s).
\label{expand:L(s)n:P}
\end{gather}
For example, since $L(s) = p_0(s,t)^{-1} P(s) -
p_0(s,t)^{-1}p_1(s,t) + u_1(s,t) + \cdots$, we have
\begin{gather}
    b^{(1)}_0(s,t) = p_0(s,t)^{-1},\qquad
    b^{(1)}_1(s,t) = - p_0(s,t)^{-1}p_1(s,t) + u_1(s,t).
\label{b(1)0,1}
\end{gather}
We def\/ine operator $B_n$ by
\begin{gather}
    B_n(s) := \bigl(L(s)^n\bigr)_{\geq 0} - \alpha b^{(n)}_n(s,t)
\nonumber\\
 \phantom{B_n(s)}{}   = \sum_{j=0}^{n-1} b^{(n)}_j(s,t) P^{(n-j)}(s)
    + (1-\alpha)   b^{(n)}_n(s,t).
\label{def:Bn(s)}
\end{gather}
Here $(\cdot)_{\geq 0}$ is the projection of a microdif\/ferential
operator to the dif\/ferential operator part.  It is easy to see that
condition \eqref{p0,u1,p1} is equivalent to $B_1(s)=\der$ and that
$B_n(s) = (L(s)^n)_{>0}$ when $\alpha=1$.

The def\/inition of the cmKP hierarchy with a gauge parameter $\alpha$
is the same as the usual one, i.e., \eqref{cmkp:dL/dt},
\eqref{cmkp:LP=PL} and \eqref{cmkp:BP=PB}.

\begin{proposition}
\label{prop:gauge-trans}
The pair $(\tilde L(s),\tilde P(s))_{s\in\Integer}$ of sequences of
differential operators
\begin{gather}
    \tilde L(s) := e^{(\alpha-\beta)\varphi(s,t)} L(s)
                    e^{-(\alpha-\beta)\varphi(s,t)},
\nonumber\\
    \tilde P(s) := e^{(\alpha-\beta)\varphi(s+1,t)} P(s)
                    e^{-(\alpha-\beta)\varphi(s,t)}
\label{def:(tildeL,tildeP)}
\end{gather}
 for $s\in\Integer$ is a solution of the system \eqref{cmkp:dL/dt},
 \eqref{cmkp:LP=PL}, \eqref{cmkp:BP=PB} with gauge parameter
 $\beta$. Here $\beta$ can be~$0$.
\end{proposition}

\begin{proof}
 Let us check the condition \eqref{p0,u1,p1} f\/irst. The operators
 $\tilde L(s)$ and $\tilde P(s)$ have the form
\begin{gather}
    \tilde L(s)
    = \der + \tilde u_1(s) + \tilde u_2(s) \der^{-1} + \cdots,
\nonumber\\
    \tilde P(s)
    = \tilde p_0(s) \der + \tilde p_1(s),
\label{tildeL,tildeP}
\end{gather}
where
\begin{gather}
    \tilde u_1(s) = u_1(s) - (\alpha-\beta) \varphi'(s,t),
\nonumber\\
    \tilde p_0(s)
    = e^{(\alpha-\beta)\bigl(\varphi(s+1,t)-\varphi(s,t)\bigr)} p_0(s),
\label{tilde(u1,p0,p1)}\\
    \tilde p_1(s)
    = e^{(\alpha-\beta)\bigl(\varphi(s+1,t)-\varphi(s,t)\bigr)}
    (p_1(s) - (\alpha-\beta)p_0(s) \varphi'(s,t)).
 \nonumber
\end{gather}
 Here $\varphi'(s,t)=\der \varphi(s,t)/\der x = \der \varphi(s,t)/\der
 t_1 = b^{(1)}_1(s,t)$. Using the explicit form \eqref{b(1)0,1}, we can
 check that $(1-\beta) \tilde p_0(s,t) \tilde u_1(s,t) + \beta \tilde
 p_1(s,t) = 0$.

 The operator $P^{(n)}(s)$ def\/ined by \eqref{def:P(n)(s)} transforms as
\begin{gather*}
    P^{(n)}(s) \mapsto \tilde P^{(n)}(s)
    := e^{ (\alpha-\beta)\varphi(s+n,t)} P^{(n)}(s)
       e^{-(\alpha-\beta)\varphi(s,t)}
\end{gather*}
 by the transformation \eqref{def:(tildeL,tildeP)}. Hence $(\tilde
 L(s))^n$ is expanded as
\begin{gather}
    (\tilde L(s))^n =
    \sum_{j=0}^\infty \tilde b^{(n)}_j(s) \tilde P^{(n-j)}(s),\label{expand:tildeL}
\\
    \tilde b^{(n)}_j(s)
    :=
    e^{(\alpha-\beta)\bigl(\varphi(s,t)-\varphi(s+n-j,t)\bigr)}
    b^{(n)}_j(s).
 \nonumber
\end{gather}
 Particularly, $\tilde b^{(n)}_n(s) = b^{(n)}_n(s)$, which implies
\begin{gather*}
    \tilde B_n(s) :=
    (\tilde L(s))^n_{\geq 0} - \beta (\tilde L(s))^n_{0}
\\
  \phantom{\tilde B_n(s)}{}  =
    e^{ (\alpha-\beta)\varphi(s,t)} B_n(s)
    e^{-(\alpha-\beta)\varphi(s,t)}
    + (\alpha-\beta) b^{(n)}_n(s).
\end{gather*}
 It remains to check \eqref{cmkp:dL/dt} and \eqref{cmkp:BP=PB} by
 straightforward computation. (Equation \eqref{cmkp:LP=PL} is obvious
 from the def\/inition of $\tilde L(s)$ and $\tilde P(s)$,
 \eqref{def:(tildeL,tildeP)}.)
\end{proof}

\begin{corollary}
\label{cor:gauge-equiv}
 The cmKP hierarchies with different gauge parameters are equivalent
 through the transformation \eqref{def:(tildeL,tildeP)}
\end{corollary}

If $\beta=0$ in \propref{prop:gauge-trans}, the resulting system is
the mKP hierarchy. \propref{prop:cmkp=mkp+gauge} is the case when
$(\alpha,\beta)=(1,0)$. See \secref{subsec:dressing-op/func}.

\section{Equivalence of two formulations}
\label{app:diff-ce=diff-l}

The two formalisms of the cmKP hierarchy discussed in
\appref{app:diff-ce} and \appref{app:cmkp:gauge} are equivalent. The
proof is almost straightforward computation but lengthy.

Rewriting the dif\/ference operator formalism to the dif\/ferential operator
formalism is essentially the same as the procedure described in \S~1.2 of
\cite{uen-tak:84}, where the KP hierarchy is embedded in the Toda
lattice hierarchy. Assume that a solution $\boldL = \sum\limits_{j=0}^\infty
b_j(s,t)e^{(1-j)\der_s}$ of the system \eqref{cmkp:diff-ce} is
given. The idea is to interpret the operator $\der_{t_1} - \boldB_1$ as
the operator $b_0(s,t) (P(s) - e^{\der_s})$. Namely, we def\/ine the
operator $P$ by
\begin{gather}
    P(s) := b_0(s,t)^{-1} (\der - (1-\alpha) b_1(s,t)),
\label{P(s)<-boldB1}
\end{gather}
where $t_1$ is replaced by $t_1+x$. (We do not write the dependence
on $x$ explicitly.) Using the operator $P^{(n)}(s)$ def\/ined by
\eqref{def:P(n)(s)}, we def\/ine the $L$ operator by
\begin{gather}
    L(s) := \sum_{j=0}^\infty b_j(s,t) P^{(1-j)}(s)
\nonumber\\
    \phantom{L(s)}{}= b_0(s,t) P(s) + b_1(s,t) + b_2(s,t) P^{(-1)}(s) + \cdots
\nonumber\\
   \phantom{L(s)}{}=
    \der + \alpha b_1(s,t) + \cdots.
\label{L(s)<-boldL}
\end{gather}
The condition \eqref{p0,u1,p1} is automatically satisf\/ied, thus we
have $B_1(s)=\der$.

First we prove \eqref{cmkp:LP=PL}. The left hand side of
\eqref{cmkp:LP=PL} is
\begin{gather}
    L(s+1) P(s) = \sum_{j=0}^\infty b_j(s+1) P^{(2-j)}(s),
\label{L(s+1)P(s)}
\end{gather}
by the def\/inition \eqref{def:P(n)(s)}. The right hand side of
\eqref{cmkp:LP=PL} is
\begin{gather}
P(s) L(s) =
    P(s) \sum_{j=0}^\infty b_j(s) P^{(1-j)}(s)
\nonumber\\
\phantom{P(s) L(s)}{}    =
    b_0(s)^{-1}
    \sum_{j=0}^\infty
    \left(\frac{\der b_j(s)}{\der x}
         + b_j(s) \der - (1-\alpha) b_j(s) b_1(s)
    \right)P^{(1-j)}(s).
\label{P(s)L(s):temp}
\end{gather}
Since $\der b_j(s)/\der x = \der b_j(s)/\der t_1$, the Lax equation
\eqref{cmkp:diff-ce} with $n=1$ gives information on $\der
b_j(s)/\der x$. Comparing the coef\/f\/icients of $e^{(1-j)\der_s}$ in
\eqref{cmkp:diff-ce}, we have
\begin{gather}
    \frac{\der b_j(s)}{\der x}
    =
    b_0(s) b_{j+1}(s+1) - b_0(s-j) b_{j+1}(s)
\nonumber\\
\phantom{\frac{\der b_j(s)}{\der x}=}{}+ (1-\alpha)b_1(s) b_j(s) - (1-\alpha)b_1(s+1-j) b_j(s).
\label{dbj(s)/dx}
\end{gather}
Substituting it into \eqref{P(s)L(s):temp} and using the property
\eqref{P(n)(s):composition} of $P^{(j)}(s)$, we have
\begin{gather}
    P(s) L(s) =
    \sum_{j=0}^\infty  b_j(s+1) P^{(2-j)}(s),
\label{P(s)L(s):final}
\end{gather}
which, together with \eqref{L(s+1)P(s)}, proves \eqref{cmkp:LP=PL}.
The following formula is a consequence of \eqref{cmkp:LP=PL}:
\begin{gather}
    L(s+m) P^{(m)}(s) = P^{(m)}(s) L(s).
\label{cmkp:LP(n)=P(n)L}
\end{gather}

The proof of \eqref{cmkp:BP=PB} is almost the same. Note that if we
expand $\boldL^n$ as
\begin{gather}
    \boldL^n = \sum_{j=0}^\infty b^{(n)}_j(s) e^{(n-j)\der_s},
\label{boldLn}
\end{gather}
operator $L(s)^n$ is expanded as \eqref{expand:L(s)n:P} with the
same coef\/f\/icients $b^{(n)}_j(s)$ as in \eqref{boldLn}. This is
proved by induction with the help of \eqref{cmkp:LP=PL} proved above
and \eqref{P(n)(s):composition}.

Hence the coef\/f\/icients $b^{(n)}_j(s)$ in \eqref{def:Bn(s)} are the
same as in \eqref{boldLn}. Using this fact and
\eqref{cmkp:diff-ce:zs} $m=1$, we can prove \eqref{cmkp:dP/dt},
i.e., \eqref{cmkp:BP=PB}. We omit details which are similar to the
above proof of~\eqref{cmkp:LP=PL}. The formula
\begin{gather}
    (\der_{t_n} - B_n(s+m)) P^{(m)}(s)
    = P^{(m)}(s) (\der_{t_n} - B_n(s)),
\label{cmkp:BP(n)=P(n)B}
\\
    \text{i.e.,}\quad 
    \frac{\der P^{(m)}(s)}{\der t_n}
    = B_n(s+m) P^{(m)}(s) - P^{(m)}(s) B_n(s),
\label{cmkp:dP(m)(s)/dtn=[B,P(m)]}
\end{gather}
derived from \eqref{cmkp:BP=PB} shall be used in the following step.

Let us proceed to the proof of the Lax equation \eqref{cmkp:dL/dt}.
Its left hand side is
\begin{gather}
    \frac{\der L(s)}{\der t_n}
    =
      \sum_{j=0}^\infty \frac{\der b_j(s)}{\der t_n} P^{(1-j)}(s)
\nonumber\\
\phantom{\frac{\der L(s)}{\der t_n}=}{} + \sum_{j=0}^\infty  \sum_{k=0}^n
       b_j(s) b^{(n)}_k(s+1-j) P^{(n+1-j-k)}(s)
    - L(s) B_n(s).
\label{dL(s)/dtn:temp1}
\end{gather}
(We used \eqref{cmkp:dP(m)(s)/dtn=[B,P(m)]} and
\eqref{P(n)(s):composition}.)  On the other hand,
\eqref{cmkp:LP(n)=P(n)L} and \eqref{P(n)(s):composition} imply
\begin{gather}
    [B_n(s), L(s)]
    =
    \sum_{j=0}^n \sum_{k=0}^\infty
    b^{(n)}_j(s) b_k(s+n-j) P^{(n+1-j-k)}(s) - L(s) B_n(s).
\label{dL(s)/dtn:temp2}
\end{gather}
In order to prove that \eqref{dL(s)/dtn:temp1} and
\eqref{dL(s)/dtn:temp2} are equal, we have only to show
\begin{gather*}
    \frac{\der b_l(s)}{\der t_n}
    =
    \sum_{\substack{0\leq j\leq n, 0\leq k \\ j+k = l+n}}
    \Bigl( b^{(n)}_j(s) b_k(s+n-j) - b_k(s) b^{(n)}_j(s+1-k) \Bigr),
\end{gather*}
which is nothing but the coef\/f\/icient of $e^{(1-l)\der_s}$ of
\eqref{cmkp:diff-ce}. Thus we have proved that a solution of the
dif\/ference operator equations \eqref{cmkp:diff-ce} gives a solution
of the system \eqref{cmkp:dL/dt}, \eqref{cmkp:LP=PL} and
\eqref{cmkp:BP=PB}.

Conversely, when a solution $(L(s), P(s))_{s\in\Integer}$ of the
system \eqref{cmkp:dL/dt}, \eqref{cmkp:LP=PL}, \eqref{cmkp:BP=PB} is
given and $L(s)$ is expanded as
\begin{gather}
    L(s) = \sum_{j=0}^\infty b_j(s) P^{(1-j)}(s),
\label{expand:L(s)}
\end{gather}
then the $\boldL$ operator def\/ined by \eqref{bL} satisf\/ies the
system \eqref{cmkp:diff-ce}. Note that, due to condition~\eqref{p0,u1,p1},
 $B_1(s)=\der$, which means that $t_1$ and $x$
always appear in the form $t_1+x$. Hence we can eliminate $x$ by
just replacing $t_1+x$ by $t_1$. The Lax equation
\eqref{cmkp:diff-ce} for the dif\/ference operator is proved by
tracing back the above proof of \eqref{cmkp:dL/dt}.

\begin{remark}
This correspondence holds also for the case $\alpha=0$. If $p_0(s)$
is normalized to $1$, the system \eqref{cmkp:dL/dt},
\eqref{cmkp:LP=PL}, \eqref{cmkp:BP=PB} is the mKP hierarchy in
\cite{dic:99,tak:02}. Its equivalence to the system
\eqref{cmkp:diff-ce}\footnote{In \cite{dic:99} the system
\eqref{cmkp:diff-ce} ($\alpha=0$) is called the ``discrete KP'', but
such a name is used more often for the system with discrete
independent variables. So we call \eqref{cmkp:diff-ce} the
``dif\/ference operator formalism'' of the cmKP hierarchy.} was proved
in \cite{dic:99} by a dif\/ferent method.
\end{remark}

\subsection*{Acknowledgments}
The work of T.~Takebe is supported in part by JSPS grant 18540210. He is
also grateful to Saburo Kakei and Kanehisa Takasaki for their interest
and comments.
The work of L.P.~Teo is supported in part by MMU funding
PR/2006/0590. She would also like to thank Academia Sinica of Taiwan
for its hospitality during her stay, when part of this work was
done.

\LastPageEnding


\begin{thebibliography}{99}
\footnotesize

\bibitem{CT}
Chang J.H.,  Tu M.H., On the {M}iura map between the
dispersionless
  {KP} and dispersionless modif\/ied {KP} hierarchies, {\it J. Math. Phys.}, 2000, V.41, 5391--5406,
   \href{http://arxiv.org/abs/solv-int/9912016}{solv-int/9912016}.


\bibitem{dic:99}
 Dickey L.A., Modif\/ied KP and discrete KP, {\it Lett. Math. Phys.}, 1999, V.48, 277--289,
 \href{http://arxiv.org/abs/solv-int/9902008}{solv-int/9902008}.

\bibitem{djkm}
 Date E., Kashiwara M., Jimbo M., Miwa T., Transformation
groups for soliton equations, in Nonlinear Integrable Systems
-- Classical Theory and Quantum Theory, Singapore, World Scientif\/ic, 1983, 39--119.


\bibitem{Duren}
 Duren P.L., Univalent functions, {\it Grundlehren der
Mathematischen Wissenschaften},
Vol.~259, New York, Springer-Verlag, 1983.

\bibitem{kas-miw:81}
 Kashiwara M., Miwa T.,
Transformation groups for soliton equations. I. The $\tau$ function of
the Kadomtsev--Petviashvili equation, {\it Proc. Japan Acad. Ser. A Math. Sci.}, 1981, V.57,
342--347.

\bibitem{kac-pet:86}
 Kac V.G., Peterson D.H., 
Lectures on the inf\/inite wedge-representation and the MKP hierarchy,
in  Syst\`emes dynamiques non lin\'eaires: int\'egrabilit\'e et comportement
qualitatif, {\it S\'em. Math. Sup.}, Vol.~102, Montreal,
Presses Univ. Montr\'eal, 1986, 141--184.

\bibitem{kup:00}
 Kupershmidt B.A.,
KP or mKP. Noncommutative mathematics of Lagrangian,
Hamiltonian, and integrable systems,
{\it Mathematical Surveys and Monographs}, Vol.~78, 
 Providence, RI,  American Mathematical Society, 2000.

\bibitem{Kuper}
 Kupershmidt B.A., The quasiclassical limit of the modif\/ied
{KP}
  hierarchy, {\it J. Phys. A: Math. Gen.}, 1990, V.23, 871--886.


\bibitem{Pom}
Pommerenke C., Univalent functions, G\"ottingen, Vandenhoeck \& Ruprecht,
   1975 (with a chapter on quadratic dif\/ferentials by Gerd Jensen,
  Studia Mathematica/Mathematische Lehrb\"ucher, Band XXV).

\bibitem{sat:81}
 Sato M., Soliton equations as dynamical systems on an inf\/inite
dimensional Grassmann manifold, {\it RIMS Kokyuroku}, 1981, V.439,
30--46.

\bibitem{sat-nou:84}
 Sato M., Noumi M., Soliton equations and the universal
Grassmann manifolds, {\it Sophia University Kokyuroku in
Math.}, Vol.~18,  Tokyo, Sophia University, 1984 (in Japanese).

\bibitem{sat-sat:82}
 Sato M., Sato Y., Soliton equations as dynamical systems on
inf\/inite dimensional Grassmann manifold, in Nonlinear Partial
Dif\/ferential Equations in Applied Science, Proceedings of the
U.S.--Japan Seminar (1982, Tokyo), {\it Lect. Notes in Num. Anal.}, 1982, V.5, 259--271.

\bibitem{taka:84}
Takasaki K., Initial value problem for the Toda lattice hierarchy,
{\it Adv. Stud. Pure Math.}, Vol.~4, 
Group Representations and Systems of Dif\/ferential Equations (1982, Tokyo),
Amsterdam, North-Holland, 1984, 139--163.

\bibitem{tak:90}
 Takebe T., Toda lattice hierarchy and conservation laws, 
{\it Comm. Math. Phys.}, 1990, V.129, 281--318.

\bibitem{tak:91a}
 Takebe T., Representation theoretical meaning of the initial value
problem for the Toda lattice hierarchy.~I, {\it Lett. Math. Phys.}, 1991, V.21, 77--84.

\bibitem{tak:91b}
Takebe T., Representation theoretical meaning of the initial value
problem for the Toda lattice hierarchy.~II, {\it Publ. RIMS}, 1991, V.27, 491--503.

\bibitem{tak:02}
 Takebe T., A note on modif\/ied KP hierarchy and its (yet another)
dispersionless limit, 
{\it Lett. Math. Phys.}, 2002, V.59, 157--172, 
\href{http://arxiv.org/abs/nlin.SI/0111012}{nlin.SI/0111012}.

\bibitem{teo:03}
 Teo L.-P., On dispersionless coupled modif\/ied KP hierarchy,
 \href{http://arxiv.org/abs/nlin.SI/0304007}{nlin.SI/0304007}.

\bibitem{Teo2}
 Teo L.-P., Analytic functions and integrable
  hierarchies -- characterization of tau functions, {\it Lett. Math. Phys.},
2003, V.64, 75--92,
\href{http://arxiv.org/abs/hep-th/0305005}{hep-th/0305005}.

\bibitem{TT2}
Takasaki K., Takebe T., ${\rm SDif\/f}(2)$ {KP} hierarchy,
in Inf\/inite
  Analysis, Part A, B (1991, Kyoto), {\it  Adv. Ser. Math. Phys.}, Vol.~16,  River Edge, NJ, World Sci.
  Publishing, 1992, 889--922, \href{http://arxiv.org/abs/hep-th/9112046}{hep-th/9112046}.

\bibitem{TT1}
Takasaki K., Takebe T., Integrable hierarchies and
dispersionless
  limit, {\it Rev. Math. Phys.}, 1995, V.7, 743--808, \href{http://arxiv.org/abs/hep-th/9405096}{hep-th/9405096}.

\bibitem{tak-tak:05}
Takasaki K., Takebe T., L\"owner equations and dispersionless
hierarchies,  in the Proceedings of XXIII International
Conference of Dif\/ferential Geometric Methods in Theoretical Physics
Nankai Institute of Mathematics (August 2005, Tianjin, China), to appear, 
\href{http://arxiv.org/abs/nlin.SI/0512008}{nlin.SI/0512008}.

\bibitem{uen-tak:84}
Ueno K., Takasaki K., Toda lattice hierarchy, 
{\it Adv. Stud. Pure Math.}, Vol.~4, 
Group Representations and Systems of Dif\/ferential Equations (1982, Tokyo), 
Amsterdam, North-Holland, 1984, 1--95.

\end{thebibliography}
\end{document}